\documentclass[12pt]{article}
\usepackage{a4}
\usepackage{array}
\usepackage{fancybox}
\usepackage{hhline}
\usepackage{amsmath,amssymb}
\usepackage{amsmath,amsfonts}
\usepackage[dvips]{graphics}
\usepackage{graphicx}
\usepackage{epsfig}
\def\bmp{\mbox{\boldmath $p$}}
\def\bmd{\mbox{\boldmath $\Delta$}}
\def\bmpart{\mbox{\boldmath $\partial$}}
\begin{document}

\pagestyle{empty}

\hfill LAL/99-06

\vspace{2cm}

\begin{center} 
{\bf \Large A NEW GLOBAL ANALYSIS OF \\
DEEP INELASTIC SCATTERING DATA \\}

\vspace{1cm}

{\large V.~Barone$^{a,b}$,
 C.~Pascaud$^{c}$ and F.~Zomer$^{c}$ \\}
\vspace{0.5cm}
 
$^a$ Dipartimento di Fisica Teorica, Universit{\`a} di Torino\\
and INFN, Sezione di Torino, via P.~Giuria 1, I-10125, 
Torino, Italy \\
$^b$ D.S.T.A., Universit{\`a} ``A.~Avogadro'', 
\\ 
c.so Borsalino 54, I-13100 Alessandria, Italy \\
$^c$ Laboratoire de l'Acc\'el\'erateur Lin\'eaire, IN2P3-CNRS \\
 and Universit\'e de Paris-Sud,
 F-91898 B.P.~34 Orsay Cedex, France
\end{center}

\vspace{1cm}

\begin{center}
{\bf \large Abstract}
\end{center}

\noindent A new QCD analysis of Deep Inelastic 
Scattering (DIS) data is presented. 
All available  neutrino and anti-neutrino cross sections 
are reanalysed and included in the fit, along with 
charged-lepton DIS  and Drell-Yan data. 
A massive factorisation scheme is used to describe
the charm component of the structure functions. 
Next-to-leading order parton distribution functions 
 are provided. 
In particular, the strange sea density is determined 
with a higher accuracy with respect to other global fits.

\newpage
\pagestyle{plain}

\section{Introduction}
\label{Introduction}

In Deep Inelastic Scattering (DIS) processes a 
neutral or charged lepton $\ell$ 
interacts with a nucleon $N$ yielding a lepton $\ell'$
 and a set of undetected hadrons $X$ in the final state.
 The kinematics of this process, 
$\ell(k)+ N(p) \rightarrow \ell'(k')+X$,  
 is determined by two independent variables, besides 
the energy of the incoming lepton. One usually
chooses them among the four Lorentz invariants
\[
Q^2 \equiv -q^2=-(k-k')^2,\;
x=\frac{Q^2}{2p\cdot q}\,,\; y=\frac{p\cdot q}{p\cdot k},\;
 W^2=(q+p)^2,
\]
which are obtained experimentally by measuring
 the momentum, the direction of the scattered lepton and the 
initial 
 momenta $k$, $p$. 

 Studies of DIS processes, both on the experimental
 \cite{exp} and on the 
 theoretical side \cite{tung-revue},
 have shed light on the nucleon internal dynamics.
 In particular, powerful tests of perturbative
 Quantum Chromo Dynamics (pQCD) have been carried out, 
consisting  
of global analyses performed on a large class of DIS observables. 
 Among these
 the structure functions measured in 
neutrino (anti-neutrino) DIS
 play a major role for the  determination of the
 flavour distributions, in particular  the valence 
and the strange sea densities. 

What  experiments directly measure is a
differential cross-section from which the structure functions 
are extracted by means of a theoretical analysis. This includes 
the application of 
electroweak radiative corrections, the 
 determination of
$R = \sigma_L/\sigma_T$, the incorporation of 
possible nuclear effects, etc~...
 In particular, 
the analysis of neutrino data is a
difficult and delicate procedure   
(see \cite{seligman} for a clear and detailed
 description), and so far  
only a small part of the information accumulated 
in $\nu (\bar \nu)$DIS has been exploited in the QCD parametrisations. 

The purpose of this paper is to present a new global analysis
of DIS data which includes the available $\nu$ and $\bar \nu$ 
 cross section measurements\footnote{This type 
of analysis was proposed long ago by M.W. Krasny \cite{witek}.}, 
besides the structure function data collected in 
charged--lepton DIS  experiments. 
We resort directly to the $\nu (\bar \nu)$ DIS differential 
cross sections,  
 avoiding to use the  
neutrino structure function results. 
The latter are the 
product of a preanalysis which may be (and often indeed is) 
based on theoretical assumptions different from 
those of the global fit that one is performing. 
Thus, for a full consistency, we use
only {\em cross section} data. This limits  
our neutrino (and anti-neutrino) data set to the 
 BEBC (Hydrogen target) \cite{bebc},
 CDHS (Hydrogen and Deuterium targets) \cite{cdhs} 
 and CDHSW (Iron target) \cite{cdhsw} measurements. 
The CHARM \cite{charm} and CCFR \cite{ccfr-data} 
experiments do not provide cross sections but only 
structure functions and hence their results are not 
included in our 
analysis.

The problem with the BEBC, CDHS and CDHSW 
data is that they cannot be used in the form they
were published, since the electroweak radiative corrections
were either incomplete or not applied at all and/or the bin centre
corrections were not performed. Thus we had to reevaluate 
the neutrino cross sections to take all these corrections 
into account. This is the preliminary step of our analysis. 
Since most of the $\nu$ ($\bar \nu$) DIS data come from nuclear targets, 
nuclear corrections must also be applied. 

Besides the neutrino data, the structure function measurements 
from charged--lepton DIS experiments (NMC \cite{nmc}, BCDMS \cite{bcdms},
 H1 \cite{h1-94}) 
 are included in our fits. 
The only non DIS data we use are the 
Drell--Yan \cite{e605,na51,e866} 
data which constrain  the light
sea. 

Our fitting procedure is designed in such a way to take 
properly into account the experimental uncertainties and the 
correlations among them, which are known to affect the 
$Q^2$ slopes
of the structure functions and ultimately the determination 
of the parton densities \cite{mimil}.

An important feature of the QCD analysis presented in this paper
is the  accurate treatment of the charm contribution 
to the structure functions. A massive factorisation scheme is used, the 
so-called Fixed Flavour Scheme \cite{tfns}. It is known \cite{io} 
that for a precise extraction of the strange sea density,  
charm mass effects cannot be neglected and have to be correctly 
incorporated.

In  extracting the 
parton distributions, 
 the neutrino data (in 
particular the high--statistics CDHSW data)
add a great quantity of information to that coming from 
charged-lepton DIS. The latter 
is insufficient to constrain all flavour distributions, 
being essentially limited to one observable, $F_2$.  
Charged--current DIS provides four more independent 
combinations of parton densities, $F_2^{\nu}$, $F_2^{\bar \nu}$, 
$x F_3^{\nu}$, $x F_3^{\bar \nu}$. 
As a consequence, the abundance of  neutrino data 
in our fit ensures an excellent 
accuracy in the determination of the flavour 
densities. 

A special emphasis will be given to the strange sea
density. Due to the lack of data able to constrain 
it, in the existing fits \cite{cteq,mrs}
this distribution is tightly related to the 
non-strange sea distributions and 
essentially borrowed\footnote{In the GRV fit \cite{grv} 
the strange distribution is assumed to be zero at a 
very low $Q^2$ scale and then entirely generated by 
the QCD evolution.} 
 from the CCFR extraction \cite{ccfr-strange}. 
Clearly, this is not a consistent procedure. 
Here we present the first fully consistent 
determination of the strange distribution 
within a global fit of all 
parton densities. The wealth of neutrino and anti-neutrino
data will also allow us to test the possible charge asymmetry 
of the strange sea ($s \ne \bar s$) predicted by some 
authors.

In the present work the strong coupling constant is 
independently fixed to a value close to the 
world average. In a forthcoming paper 
we shall study the possibility of determining $\alpha_s$ 
from the minimisation of the total $\chi^2$ of the fit, 
and discuss the stability of this determination and its 
correlation with the gluon density.

The article is organised as follows. In 
section~\ref{QCD-phenomenology} we collect the main theoretical 
ingredients of the QCD analysis of DIS data. 
The re-evaluation of the neutrino and anti-neutrino differential cross
sections is the content of section~\ref{Re-evaluation}.
The fitting procedure is described in section~\ref{qcd-analysis}. 
Finally, the results 
on  cross sections, 
structure functions and parton distribution functions are 
presented in section~\ref{results}.

\section{Deep inelastic scattering in QCD}
\label{QCD-phenomenology}
The differential cross section of neutral--current (NC) DIS
of charged leptons ($\ell$), in the one--boson 
approximation and for moderate $Q^2\ll M_Z^2$, is given by
\begin{equation}\label{nc-section-formule}  
\frac{d^2\sigma^{\ell N}}{dxdy}=
\frac{8\pi\alpha^2_{em}M_NE}{Q^2}
\biggl[
xy^2 \, F_1^{\ell N}(x,Q^2)+
\biggl(1-y-\frac{M_Nxy}{2E}\biggr)F_2^{\ell N}(x,Q^2)
\biggr],
\end{equation}
If $Q^2\ll M_Z^2$ charged--lepton NC DIS is 
essentially an electromagnetic reaction (that 
is dominated by one--photon exchange), the $Z^0$ 
contribution being totally negligible. 
In eq.~(\ref{nc-section-formule}) $\alpha_{em}=1/137$ is 
the electromagnetic coupling constant, 
$E$ is the beam energy and  $M_N$ is the nucleon mass. 

For charged--current (CC) neutrino (anti-neutrino) DIS
one has
\begin{eqnarray}\label{cc-section-formule}  
\frac{d^2\sigma^{\nu(\bar{\nu})N}}{dxdy}&=&
\frac{G_F^2M_NE}{\pi}\biggl(\frac{M_W^2}{Q^2+M_W^2}\biggr)^2\nonumber\\
&\times& \biggl[
xy^2\, F_1^{\nu(\bar{\nu})N}(x,Q^2)+
\biggl(1-y-\frac{M_Nxy}{2E}\biggr)F_2^{\nu(\bar{\nu})N}(x,Q^2)
\nonumber\\
&+(-)& \biggl(y-\frac{y^2}{2}\biggr)xF_3^{\nu(\bar{\nu})N}(x,Q^2)
\biggr],
\end{eqnarray}
where 
 $M_W$ is $W$--boson mass and  
$G_F$ is the Fermi coupling
constant.

\begin{table}[htbp]
  \begin{center}
    \leavevmode
\begin{small}
    \begin{tabular}{|c|c|c|}
\hline
& &  \\
          &$F_{2,l}$      & $F_{2,c}$   \\\hline
& &  \\
$\ell^\pm p$ &$x[\frac{4}{9}(u+\bar{u})+\frac{1}{9}(d+\bar{d}+s+\bar{s})]$&
           $\frac{4}{9} \, (\frac{\alpha_s}{2 \pi})\,  
             C^{c,(0)}_{2,g}\otimes xg$ 
\\ & &             \\\hline & &  \\
$\ell^\pm n$  &$x[\frac{4}{9}(d+\bar{d})+\frac{1}{9}(u+\bar{u}+s+\bar{s})]$
 &
           $\frac{4}{9} \, (\frac{\alpha_s}{2 \pi})\,  
            C^{c,(0)}_{2,g}\otimes xg$ 
\\ & &            \\\hline  & &  \\ 
$\nu p$   & $2x[\bar{u}+d|V_{ud}|^2+s|V_{us}|^2]$&
          $2\xi[d(\xi)|V_{cd}|^2+s(\xi)|V_{cs}|^2]$
\\ & &             \\\hline & &  \\
$\nu n$   & $2x[\bar{d}+u|V_{ud}|^2+s|V_{us}|^2]$&
          $2\xi[u(\xi)|V_{cd}|^2+s(\xi)|V_{cs}|^2]$
\\& &             \\\hline & &  \\
$\bar{\nu} p$  & $2x[u+\bar{d}|V_{ud}|^2 +\bar{s}|V_{us}|^2 ]$&
          $2\xi[ \bar{d}(\xi)|V_{cd}|^2+ \bar{s}(\xi)|V_{cs}|^2]$
\\ & &           \\\hline & &  \\  
$\bar{\nu} n$  &  $2x[d+\bar{u}|V_{ud}|^2 +\bar{s}|V_{us}|^2] $&
          $2\xi [\bar{u}(\xi)|V_{cd}|^2+ \bar{s}(\xi)|V_{cs}|^2]$
\\ & &            \\\hline 

    \end{tabular}
\end{small}
    \caption{\small Leading--order expressions of 
$F_2$ ($l$: light sector; $c$: charm sector in the FFS).
 The $x$ and $Q^2$ arguments have been omitted. Only the 
slow-rescaled 
argument $\xi$ has been explicitly indicated.}
    \label{tab:str-f2}
  \end{center}
\end{table}

\begin{table}[htbp]
  \begin{center}
    \leavevmode
\begin{small}
    \begin{tabular}{|c|c|c|}
\hline
& &  \\
          &$xF_{3,l}$      & $xF_{3,c}$   \\\hline & &  \\
$\nu p$   & $2x[d|V_{ud}|^2+s|V_{us}|^2-\bar{u}]$&
          $2\xi[d(\xi)|V_{cd}|^2+s(\xi)|V_{cs}|^2]$
\\ & &           \\\hline & &  \\
$\nu n$   & $2x[u|V_{ud}|^2+s|V_{us}|^2-\bar{d}]$&
          $2\xi[u(\xi)|V_{cd}|^2+s(\xi)|V_{cs}|^2]$
\\ & &          \\\hline & &  \\
$\bar{\nu} p$  & $2x[u-\bar{d}|V_{ud}|^2 -\bar{s}|V_{us}|^2 ]$&
          $-2\xi[ \bar{d}(\xi)|V_{cd}|^2+ \bar{s}(\xi)|V_{cs}|^2]$
\\ & &          \\\hline & &  \\  
$\bar{\nu} n$  &  $2x[d-\bar{u}|V_{ud}|^2 -\bar{s}|V_{us}|^2] $&
          $-2\xi [\bar{u}(\xi)|V_{cd}|^2+ \bar{s}(\xi)|V_{cs}|^2]$
  \\ & &       \\\hline 

    \end{tabular}
\end{small}

    \caption {\small Same as table \ref{tab:str-f2}, for $xF_3$.}
    \label{tab:str-f3}
  \end{center}
\end{table}

In QCD the structure functions $F_2$, $F_L = F_2 - 2x F_1$ 
and $x F_3$ are given by the 
convolution of the 
 parton distribution functions (pdf's) with some perturbatively 
calculable coefficient functions. 
In the kinematic range covered by the analysis presented in 
this article the contribution of the $b$ quark is 
negligible and the only heavy quark considered
is charm. Therefore we split the structure functions 
in two components
\[
F_i(x,Q^2)=F_{i,l}(x,Q^2)+F_{i,c}(x,Q^2), \quad i=2,3,L
\]
where $F_{i,l}(x,Q^2)$ is the light-parton contribution and 
$F_{i,c}(x,Q^2)$ is the charm contribution. In the CC case 
the latter mixes charm with light quarks. 

The massive scheme that we adopt  is the Fixed Flavour
Scheme (FFS) \cite{tfns} in which charm is a `heavy' quark in 
absolute sense.  This means that there is 
no such thing as the charm density function and charm 
is radiatively produced. Consequently
the number of active flavours is set 
to 3 irrespective of $Q^2$. The FFS has been shown to 
be more stable than the alternative massive scheme, the 
Variable Flavour Scheme (VFS) \cite{tung}, 
at moderate $Q^2$ where most of the neutrino data lie
\cite{io2,io3}. 

In the strong coupling, 
 heavy quark thresholds are accounted for according to the 
prescription of Ref.~\cite{marciano} (with $m_c=1.5$ GeV and
 $m_b=4$ GeV). 

The leading--order (LO) expressions for $F_2$ and 
$xF_3$ are collected in Tables
\ref{tab:str-f2} and \ref{tab:str-f3}. Note that:  

\begin{itemize}
\item
$\xi=x(1+Q^2/m_c^2)$
 is the slow-rescaling variable; $m_c=1.5$ GeV is the charm mass.
\item 
$V_{ij}$ are the Cabibbo-Kobayashi-Maskawa 
matrix elements.  We shall use
 $|V_{us}|=|V_{cd}|=0.224$ and $|V_{ud}|=|V_{cs}|=\sqrt{1-|V_{us}|^2}$.
\item
The symbol $\otimes$ stands for  convolution: 
\[ f\otimes g=\int_{ax}^1 \frac{dz}{z}\,  f(z) g(x/z)  \,, \]
where $a =1$ for the light sector, $a = 1 + 4 m_c^2/Q^2$
for NC charm production, $a = 1 + m_c^2/Q^2$ for 
CC charm production.  
\item
The LO charm contribution to $F_2$ in the neutral-current
charged-lepton 
case is a ${\cal O}(\alpha_s)$ quantity in the Fixed Flavour Scheme. 
$C^{c,(0)}_{2,g}$ is the LO Wilson coefficient for 
the photon--gluon fusion process \cite{witten}. Explicitly:
\begin{equation}
F_{2,c}^{\ell N}(x,Q^2) = \frac{4}{9}  \, 
(\frac{\alpha_s}{2 \pi}) \,   
C^{c,(0)}_{2,g}(m_c^2/Q^2) \otimes x g(\mu^2) \,,
\label{f2clo}
\end{equation}
where the  strong coupling is evaluated at 
the factorisation scale $\mu$, and  
$C^{c,(0)}_{2,g}(z, m_c^2/Q^2)$ can be found 
for instance in \cite{tfns}. 
\item
The charm production is different in neutral and charged current 
DIS.  In the former case it is given at LO by gluon splitting 
into a $c \bar c$ pair. 
In the latter case it is given by the direct
process $W^{+}s \rightarrow c$, with the slow-rescaling
variable taking into account the effect of the charm mass. 
\item
At order ${\alpha_s^0}$ 
the longitudinal structure function $F_L = 
F_2 - 2x F_1$ vanishes. 
\end{itemize}

The QCD analysis performed in this paper is at the  
next-to-leading order (NLO) level (the renormalization scheme 
adopted is $\overline{MS}$). 
NLO means ${\cal O}(\alpha_s)$ for the light sector
and for the charm contribution to CC DIS, 
${\cal O}(\alpha_s^2)$ for the charm contribution to NC DIS
in the Fixed Flavour Scheme. 
Since 
$F_L$ vanishes at order $\alpha_s^0$,
for consistency 
with the treatment of the charm structure 
function, 
we include in our NLO analysis the order $\alpha_s^2$
contribution to $F_L$, except for the strange-charm 
component of  $F_L$, for which the ${\cal O}(\alpha_s^2)$
longitudinal Wilson coefficients are not known yet. 
Anyway, this contribution
has a very little effect in the kinematic domain of 
our analysis. 

The light-parton 
components of the structure functions have the form
(for illustration we write only $F_2$ for the 
electromagnetic case):
\begin{equation}\label{qdis-nlo}
F_{2,l}^{\ell N}(x,Q^2) = \sum_{f=q, \bar q} e_f^2  
\{(1+\frac{\alpha_s}{2\pi} \, C_{2,f}^{(1)})\otimes xf +
\frac{\alpha_s}{2\pi} \, C_{2,g}^{(1)}\otimes xg \}\,.
\end{equation}
 The $\overline{MS}$ 
  Wilson coefficients $C_{i,q}^{(1)}$ and $C_{i,g}^{(1)}$ 
with $i=2,3,L$ can be found in Appendix I of Ref.~\cite{furmanski}.
As mentioned above, for $F_{L,l}$ we consider also the 
${\cal O}(\alpha_s^2)$ contributions to the 
coefficient functions calculated in 
 Ref.~\cite{nerveen-nnlo-fl-tout}\footnote{We
thank W. van Neerven for having
provided us the code which computes the order $\alpha_s^2$ Wilson
coefficients of $F_L$.}. 

The parton densities $xf$ and $xg$
are obtained 
by solving the DGLAP equations at NLO \cite{dglap}.  
For the light sector we choose $\sqrt{Q^2}$ as the factorisation 
and renormalisation scale.

The NLO expression of the 
charm contribution to the NC structure functions is
\begin{eqnarray}
F_{2,c}^{\ell N}(x,Q^2) &=&   
\frac{\alpha_s}{2 \pi} \, \left \{ 
\frac{4}{9} \, \left [ 
C^{c,(0)}_{2,g}(m_c^2/Q^2) + \frac{\alpha_s}{2 \pi}\, 
C^{c,(1)}_{2,g}(m_c^2/Q^2) \right ] 
\otimes x g(\mu^2) 
\nonumber \right.\\
&+&  
\frac{\alpha_s^2}{(2 \pi)^2} \, 
\sum_{f=q, \bar q} \left [
\frac{4}{9} \, C^{c,(1)}_{2,f}(m_c^2/Q^2) \otimes 
x f(\mu^2) \right.
 \nonumber \\
&+& \left. \left.
e_f^2 \, D^{c,(1)}_{2,f}(m_c^2/Q^2)
\otimes xf(\mu^2)  \right ] \right \}\,.
\label{f2cnlo}
\end{eqnarray}
The NLO coefficients $C^{c,(1)}$ and $D^{c, (1)}$ 
 for $F_{2,c}^{\ell N}$ and $F_{L,c}^{\ell N}$ 
 have been computed in the Fixed Flavour Scheme 
 in \cite{riemersma-calc}. For our calculations 
we have used the tables presented  
in \cite{riemersma-table}. 

The NLO strange-charm component of the structure functions 
is given by (omitting the Cabibbo--suppressed term
and again writing only $F_2$ for simplicity)
\begin{equation}
F_{2,c}^{\nu N}(x,Q^2) = 2 \, 
\{(1+\frac{\alpha_s}{2\pi} H_{2,f}^{c,(1)})\otimes \xi s +
\frac{\alpha_s}{2\pi} H_{2,g}^{c,(1)}\otimes \xi g \} \,,
\label{f2scnlo}
\end{equation}
The Wilson coefficients $H_{i,f}^{c,(1)}$ and 
$H_{i,g}^{c,(1)}$ $(i = 2, 3, L)$
have been computed in \cite{gottschalk} and can be found, 
with a convention update,  
also in \cite{umberto,gkr}. 
The factorisation and renormalisation scale for the charm 
structure functions is chosen to be $\sqrt{Q^2 + m_c^2}$.

\section{Re-evaluation of neutrino cross sections}
\label{Re-evaluation}

\subsection{Bin centre and radiative corrections}
\label{bin}

 We 
start with a general description of the procedure 
of bin centre
 and electroweak radiative corrections. 
We shall then give the details
of the application of this procedure to the various 
data sets. 

The neutrino DIS cross-sections are determined experimentally in
bins of three kinematic variables, say $(x,y,E)$.
 The total cross section $\sigma_{ijk}^{tot}$
 corresponding to the bin $[x_i,x_{i+1}]$, $[y_j,y_{j+1}]$ 
and $[E_k,E_{k+1}]$
is given by 
\begin{eqnarray}\label{cencore}
\sigma_{ijk}^{tot}&\equiv& 
\int_{x_i}^{x_{i+1}} \! \int_{y_j}^{y_{j+1}}
\!\int_{E_k}^{E_{k+1}}
\frac{d\phi}{dE}\, \frac{d^2\sigma}{dx dy}\, dE\, dx\, dy \nonumber \\
&=& \frac{N_{ijk}}{C}
\end{eqnarray}
where $C$ is the number of scattering centres, 
$N_{ijk}$ is the number of 
events  corrected for detector effects and 
background contamination observed
in the bin $(i,j,k)$,  $d\phi/dE$ is the neutrino beam energy flux.

 In order to relate 
$\sigma_{ijk}^{tot}$ to the differential cross section, we 
invoke the 
average theorem: there exists at 
least one point $(\bar{x}_i,\bar{y}_j,\bar{E}_k)$
inside the bin $(i,j,k)$ such that the following relation holds
\begin{equation}\label{sigma}
\frac{d^2\sigma(\bar{x}_i,\bar{y}_j,\bar{E}_k)}{dxdy}=
\frac{\sigma_{ijk}^{tot}}{S_{ijk}}
\end{equation}
where ${S_{ijk}=\int_{x_i}^{x_{i+1}}\int_{y_j}^{y_{j+1}}
\int_{E_k}^{E_{k+1}}
\frac{d\phi}{dE}\, dE\, dx\, dy}$ is the bin surface.

The differential cross section 
defined in eq.~(\ref{sigma}) must be corrected for 
electroweak radiation effects and translated to the bin centres 
$(x^c_i,y^c_j,E^c_k)$. This is done by constructing an
``experimental'' Born 
differential cross section defined as
\begin{eqnarray}\label{encoreune}
\frac{d^2\sigma^{B}_{exp}(x^c_i,y^c_j,E^c_k)}{dxdy}&\equiv&  
\frac{d^2\sigma(\bar{x}_i,\bar{y}_j,\bar{E}_k)}{dxdy} \nonumber \\
&\times&
\frac{S_{ijk}\, d^2\tilde{\sigma}^{B+R}(x^c_i,y^c_j,E^c_k)/dxdy}
{\int_{x_i}^{x_{i+1}}\int_{y_j}^{y_{j+1}}\int_{E_k}^{E_{k+1}}
\frac{d\phi}{dE}\frac{d^2\tilde{\sigma}^{B+R}}{dxdy}dE\, dx\, dy}
 \nonumber\\
&\times&
\frac{d^2\tilde{\sigma}^{B}(x^c_i,y^c_j,E^c_k)/dxdy}
{d^2\tilde{\sigma}^{B+R}(x^c_i,y^c_j,E^c_k)/dxdy}
\end{eqnarray}
where the tilde symbol designates the quantities which are
theoretically  computed, and  
 the upper-scripts $B+R$ and $B$ mean that these
 calculations are 
performed including or not, respectively, 
 the higher--order electroweak corrections. 
 The first term on the r.h.s. of 
eq.~(\ref{encoreune}) is given by
 eq.~(\ref{sigma}). The second term embodies the 
{\em bin centre corrections}
and requires a smooth parametrisation of the data. 
The third term incorporates the {\em radiative corrections}. 

 Combining the last two terms terms we get 
\begin{equation}
 \label{encoreune-bis}
  \frac{d\sigma^{B}_{exp}(x^c_i,y^c_j,E^c_k)}{dxdy}\equiv
   \frac{N_{ijk}}{C}\times
  \frac{
  d\tilde{\sigma}^{B}(x^c_i,y^c_j,E^c_k)/dxdy}{
  \int_{x_i}^{x_{i+1}}\int_{y_j}^{y_{j+1}}\int_{E_k}^{E_{k+1}}
  \frac{d\phi}{dE}\frac{d\tilde{\sigma}^{B+R}(x,y,E)}{dxdy}dE\,dx\,dy}\, .
\end{equation}

The electroweak corrections to charged current DIS were 
computed in \cite{derujula,bardin}. 
In our analysis we used the program of 
Ref.~\cite{bardin_code}, based on the results of \cite{bardin}.
At the lowest order radiative corrections include 
 the radiation of virtual and real photons from the charged lepton and
 quark legs,  and the $\gamma-W$ box diagram.
We found that these 
corrections can reach $\sim 20\%$ in some kinematic domains. 
The correction factor -- the third term in  eq.~(\ref{encoreune}) 
-- is a ratio, hence it is rather 
insensitive to  QCD corrections \cite{seligman}. Thus, for 
simplicity, we computed it at leading order in QCD. 

We also evaluated the effects of the higher--order 
$\gamma$
 radiation from the charged lepton leg \cite{blumlein_nnla}, using
the program HECTOR \cite{hector}. This correction typically
does not exceed $\sim 0.5\%$ and was applied only to 
the CDHSW data which are statistically more significant.

\subsubsection{CDHSW}\label{CDHSW-measurements}

 The CDHSW Collaboration published  \cite{cdhsw}
the $\nu Fe$ and 
 $\bar{\nu} Fe$ differential cross sections corrected for detector 
effects and background subtraction. The measurements are binned in $x$
 and $y$ for 9 different values of the neutrino beam energies
 $E^c_k$ between 23 GeV and 187.6 GeV. 
 Neither the bin centre correction in $x$
 and $y$ nor the radiative corrections were applied. Thus  
 the full 
 correction of eq.~(\ref{encoreune}) is needed. 

To evaluate the correction factors of (\ref{encoreune})  
we cannot use any of the existing
parton fits, as they do not account for nuclear effects.
Thus we adopt an iterative procedure. 
In the first step, 
 the bin centre correction and the radiative 
corrections are determined independently: 
the former by a 
 parametrisation obtained in two different ways (see below); 
the latter by a standard 
fit (we use the LO GRV pdf's \cite{grv}). 
Then, we perform a LO QCD fit to the corrected CDHSW cross 
sections, to the CDHS data and to the BCDMS, NMC and 
SLAC \cite{slac} structure functions\footnote{Nuclear corrections to 
neutrino data are applied 
as explained in section~\ref{Nuclear-targets}.}.
 Using the pdf's of this fit we 
re-evaluate the full correction factor of eq.~(\ref{encoreune})
and we iterate this step until 
the corrected differential
 cross sections get stable. In practice, to achieve the 
stability only two iterations are required. 

As we have mentioned, 
the smooth parametrisation of the data 
(required in the bin centre 
correction) is obtained by two  different methods: {\it i)}  
by a fit of the CDHSW cross sections ({\it fitting method}), 
 and {\it ii)} by an 
unfolding procedure ({\it unfolding method}). 

In the first method 
the published CDHSW $\nu Fe$ and $\bar{\nu} Fe$ differential
 cross section data are fitted to 
\[
\int_{x_i}^{x_{i+1}}\int_{y_j}^{y_{j+1}}
 \frac{d^2\tilde{\sigma}_{\nu(\bar{\nu})}(x,y,E^c_k)}{dxdy}dxdy,\quad 
k=1,...9
\]
where $\tilde{\sigma}_{\nu(\bar{\nu})}$ is computed at LO 
using simple 
Buras--Gaemers-type pdf's \cite{buras-gaemers} 
(which incorporate analytically the 
$Q^2$ dependence). 
In parallel we adopted also the unfolding method 
(described in Appendix I). 
The difference of the results obtained by the two 
procedures can be taken as a (partial) estimate of the 
uncertainty on the correction factor applied to 
neutrino cross sections. We found that 
 after two iterations the results of the two 
methods 
are compatible within 
$1\%$. 

\begin{figure}[htbp]
  \begin{center}
    \leavevmode
    \includegraphics*[width=6in]{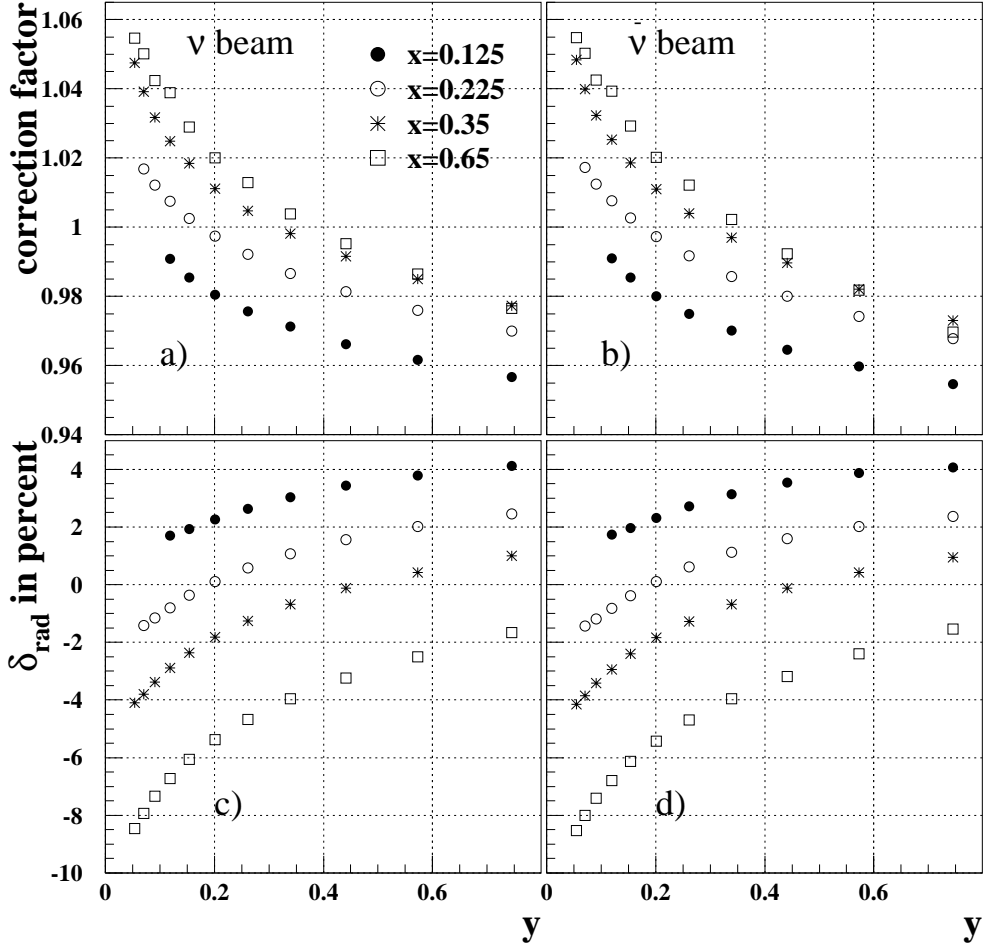}
    \caption{a) Correction factor applied to CDHSW $\nu Fe$ data 
of the 111 GeV beam energy sample, as a function
of $y$ for various $x$ bins; b) same as a) but 
for CDHSW $\bar{\nu} Fe$ data;
 c)  electroweak radiative correction factor 
 $\delta_{rad}\equiv 
(d^2\tilde{\sigma}^{B+R}/dxdy)/(d^2\tilde{\sigma}^{B}/dxdy)-1$
in percent for CDHSW $\nu Fe$; d) same as c) but for 
CDHSW $\bar{\nu} Fe$.}
    \label{cdhsw-corr}
  \end{center}
\end{figure}

 In fig.~\ref{cdhsw-corr}a,b  we show the total correction factors 
 applied to the data of the 111 GeV beam sample, as a function of $y_j$
 for fixed $x_i$. They vary between $+6\%$ and $-4\%$ and
are roughly identical for 
 neutrino and anti-neutrino beams. 
In  fig.~\ref{cdhsw-corr}c,d we show the contribution of the 
electroweak radiative corrections alone, which turns out to be the 
dominant one.

As for the normalisation of the CDHSW data, 
in \cite{cdhsw} they were normalised 
using the average total cross 
sections of \cite{sigmae}, 
namely the ratios
 $\sigma^{\nu Fe}/E_{\nu}$ ($\sigma^{\bar{\nu} Fe}/E_{\bar{\nu}}$)
 were assumed to be independent of the beam energy.
 This is a strong assumption in view of the fact that 
a linear rise with
 the energy is not experimentally excluded \cite{sigmae}. 
It is therefore important to
 check the energy dependence\footnote{We thank M.W. Krasny 
for having suggested 
to take 
 the energy dependence of the CDHSW data 
into account in our analysis.}.

\begin{table}[thbp]
  \begin{center}
    \leavevmode
    \begin{tabular}{|c|c|c|c|c|c|c|}
\hline
   \small{$\bar{a}^\nu_1$} & \small{ $\bar{a}^\nu_2$} &
\small{ $\bar{a}^{\bar{\nu}}_1$} &
\small{ $\bar{a}^{\bar{\nu}}_2$} \\
 \small{$(3\pm 5)\cdot 10^{-4}$ }&
\small{ $0.98\pm 0.05$ }& \small{$(-2.4\pm 0.5)\cdot 10^{-5}$ }
 &\small{ $1.00\pm 0.04$ }\\\hline
\end{tabular}
    \label{table-sigmae}
\caption{\small Results of the minimisation of $\chi^2_{\nu}$
 and $\chi^2_{\bar{\nu}}$. The linear functions of the fits are
defined as 
 $\sigma/E=<\sigma/E>(\bar{a}_1 E+\bar{a}_2)$ where $<\sigma/E>$ is
the average value (see text). Errors on the parameters are also given.}
  \end{center}
\end{table}

To this end we performed a linear fit
 to the measurements $\sigma^{\nu Fe}(E_\nu)/E_{\nu}$ and
  $\sigma^{\bar{\nu}Fe}(E_{\bar{\nu}})/E_{\bar{\nu}}$ of \cite{sigmae}.
 The results, renormalised by the average values \\
 $<\sigma^{\nu Fe}/E_\nu>=0.703\times 10^{-38}$cm$^2$/GeV 
and $<\sigma^{\bar{\nu} Fe}/E_{\bar{\nu}}>=0.331\times 
10^{-38}$cm$^2$/GeV,
 are given in Table~3 and shown in 
fig.~\ref{fig-sigmae} together with the 
one-standard-deviation  band, 
 computed according to the formulae of Appendix II
(we call $\chi^2_{\nu}$ and $\chi^2_{\bar \nu}$ the $\chi^2$'s 
of these fits). 
 A linear
rise of $\sigma^{\nu Fe}(E_\nu)/E_{\nu}$ with $E_{\nu}$ 
 is clearly compatible with the data.

 To take into account the effect of the uncertainty of this fit 
on the CDHSW data,
we allow the parameters $a_{\nu}$ and $a_{\bar \nu}$ to vary 
during the global pQCD fit by adding the term 
\begin{equation}\label{anuanubar}
\sum_{i,j=1}^2 
(a^\nu_i-\bar{a}^\nu_i)M_{ij}^{\nu} (a^\nu_j-\bar{a}^\nu_j)  
+(a^{\bar{\nu}}_i-\bar{a}^{\bar{\nu}}_i)M_{ij}^{\bar{\nu}}
 (a^{\bar{\nu}}_j-\bar{a}^{\bar{\nu}}_j)
\end{equation}
to the global $\chi^2$ expression. Here
$\bar{a}^\nu_i$ and $\bar{a}^{\bar{\nu}}_i$ are the parameter values
obtained from the preliminary linear fits and the matrices 
 $M_{ij}^{\nu}$ and  $M_{ij}^{\bar \nu}$ are defined as: 
 $M_{ij}^{\nu}= (1/2) \, \partial^2 \chi^2_{\nu}/\partial a^{\nu}_i
\partial a^{\nu}_j$,    
 $M_{ij}^{\bar{\nu}}=(1/2) \, 
\partial^2\chi^2_{\bar{\nu}}/\partial a^{\bar{\nu}}_i
\partial a^{\bar{\nu}}_j$. 

 Another  work done on the CDHSW data is the separation of
 correlated and uncorrelated systematic errors. 
 The correlated systematic errors are
dominated \cite{vallage} by a possible shift of the hadronic energy by
 $\pm 0.5$ GeV. As no information has been 
published, we had to estimate 
the effects of this uncertainty on the
 cross-section measurements. In CDHSW, the beam energy
 $E_{\nu}$ ($E_{\bar{\nu}}$)
 is experimentally reconstructed \cite{cdhsw} via
 $E_\nu=E_\mu+E_{X}-M_N$, where $E_\mu$ is the measured 
outgoing muon energy and 
$E_{X}$ is the measured hadronic energy in the final state.
Thus a shift of the 
hadronic energy
 induces a variation of the kinematic variables: 
 $(x,y,Q^2)\rightarrow(x_\pm,y_\pm,Q^2_\pm) $. Our estimate of the
relative differential
cross section error $\delta^\pm_{ijk}$ induced by such shifts
 is therefore
\[
\delta^\pm_{ijk}=  \frac{
\int_{x_i}^{x_{i+1}}\int_{y_j}^{y_{j+1}}
\frac{d^2\tilde \sigma^{B+R}(x^\pm,y^\pm,E^\pm_k)}{dxdy}dxdy
}{
\int_{x_i}^{x_{i+1}}\int_{y_j}^{y_{j+1}}
 \frac{d^2\tilde \sigma^{B+R}(x,y,E)}{dxdy}dxdy
}
\]
where the cross sections $\sigma^{B+R}$ are computed 
as described above.
The uncorrelated systematic errors are then obtained 
by subtracting quadratically the estimated correlated errors
from the published systematic errors
$\sigma_{ijk}^{syst}$ 
\[
 \sigma_{ijk}^{uncor}=\sqrt{(\sigma_{ijk}^{syst})^2
-{\rm max}(\delta^+_{ijk},\delta^-_{ijk})^2}. 
\]
The treatment
of both types of errors in the fit is described in 
section~\ref{qcd-analysis}.

Finally, as CDHSW  belongs
 to an old generation of experiments, 
one may worry about  possible 
sources of errors at high $y$ neglected in their analysis. 
In particular, one may question two points of the CDHSW analysis 
(see \cite{seligman} for more details):
{\it i)} 
a rejection cut against the dimuon events was used but 
no correction for it was applied; {\it ii)} 
the background coming from muon production in hadronic
 showers was not  taken into account. Using the 
recent CCFR results on  
 $\sigma^{\nu,\bar \nu}_{2\mu}/\sigma^{\nu,\bar \nu}_{1\mu}$
 \cite{ccfr-2mu} and on the production rate of muons
 in hadronic showers \cite{ccfr-hadronic}, we estimated the    
 effect of these two sources of errors on the 
measurements and found it to 
 be at the level of $1\%$ at most.

\subsubsection{CDHS}\label{cdhs-section}

 The CDHS Collaboration measured \cite{cdhs} the $\nu(\bar{\nu}) H$ 
 and $\nu(\bar{\nu}) Fe$  differential cross sections. 
Both the bin centre corrections and the 
radiative corrections were applied. The latter, however, 
are incomplete. Only  
 the $\gamma$
 radiation from the muon leg  was in fact considered. 
Hence our first step was to uncorrect back the published CDHS 
cross sections as follows
\begin{equation}
  \frac{d^2\sigma_{uncorr}(x^c_i,y^c_j,E^c_k)}{dxdy}= 
 \frac{d^2\sigma_{publ}(x^c_i,y^c_j,E^c_k)}{dxdy}
 \cdot 
 \frac{d^2\tilde{\sigma}^{B+\mu}(x^c_i,y^c_j,E^c_k)/dxdy}
{d^2\tilde{\sigma}^{B}(x^c_i,y^c_j,E^c_k)/dxdy}\;,    
\label{cdhs-uncorr}
\end{equation} 
where the  upper-script $B+\mu$ means that only
 the radiation from the muon leg was 
 included (to compute 
 it we used a modified leading--order version
of the HECTOR program). 

Then we applied to $d^2 \sigma_{uncorr}/dx \, dy$
the full radiative correction factor (i.e. the third term in 
eq.~(\ref{encoreune})
\begin{equation}
 \label{cross-section-cdhs}
  \frac{d^2\sigma^{B}_{exp}(x^c_i,y^c_j,E^c_k)}{dxdy}= 
 \frac{d^2\sigma_{uncorr}(x^c_i,y^c_j,E^c_k)}{dxdy}
 \, \frac{d^2\tilde{\sigma}^{B}(x^c_i,y^c_j,E^c_k)/dxdy}
{d^2\tilde{\sigma}^{B+R}(x^c_i,y^c_j,E^c_k)/dxdy}\;.   
\end{equation} 
 
 In the original paper \cite{cdhs} Iron and Hydrogen data 
 were normalised 
using $\sigma^{\nu Fe}/E_\nu=0.625\times 10^{-38}$cm$^2$/GeV 
and $\sigma^{\bar{\nu} Fe}/E_{\bar{\nu}}=0.3\times 10^{-38}$cm$^2$/GeV.
 Since that time, the total $\nu Fe$ and $\bar{\nu} Fe$
cross sections have been measured more precisely \cite{sigmae}.  
  We therefore applied a new overall correction factor corresponding to  
 $\sigma^{\nu Fe}/E_\nu=0.703\times 10^{-38}$cm$^2$/GeV 
and $\sigma^{\bar{\nu} Fe}/E_{\bar{\nu}}=0.331\times 
10^{-38}$cm$^2$/GeV \cite{sigmae}. 
 We have also improved the overall normalisation of the Hydrogen data 
combining the CDHS \cite{cdhs} 
$\sigma^{\nu (\bar{\nu})H}/\sigma^{\nu (\bar{\nu})Fe}$ and 
BEBC \cite{wa21} $\sigma^{\nu (\bar{\nu})H}/\sigma^{\nu (\bar{\nu})Ne}$
  results. Assuming no correlations between these experiments, 
we determined
the absolute normalisation of the CDHS Hydrogen data using: 
 $\sigma^{\nu H}/E_\nu=0.451\times 10^{-38}$cm$^2$/GeV 
and $\sigma^{\bar{\nu} H}/E_{\bar{\nu}}=0.473\times 10^{-38}$cm$^2$/GeV.
 The remaining error is $3.3\%$ for the neutrino beam and $5.3\%$ for the
 anti-neutrino beam.
  In summary, the $\nu Fe$, $\bar{\nu} Fe$, $\nu H$ and $\bar{\nu} H$
 published CDHS data have been renormalised by $+12.5\%$, $+10.3\%$,
 $+14.5\%$ and $+20.6\%$ respectively.

\subsubsection{BEBC measurements}\label{BEBC-measurements}

 Among the BEBC publications 
 only one \cite{bebc}
gives enough information to reconstruct the differential cross sections
without any QCD assumptions. In this article, the
corrected  number of events is given in bins of $x$ and $y$ 
for a given range
of $Q^2$. 
 As in the case of the CDHS data, the radiative corrections
were applied incompletely, considering only the radiation 
from the muon leg. Hence we uncorrected the published BEBC data 
analogously to what we did for CDHS (except that now we have {\em number 
of events} instead o {\em differential cross sections})
\begin{equation}
N_{ijk}^{uncorr}=
N_{ijk}^{publ} \cdot 
 \frac{
\int_{x_i}^{x_{i+1}}\int_{y_j}^{y_{j+1}}\int_{E_{min}}^{E_{max}}
\frac{d\phi}{dE}
\frac{d^2\tilde{\sigma}^{B+\mu}(x,y,E)}{dxdy}dxdydE
}
{
\int_{x_i}^{x_{i+1}}\int_{y_j}^{y_{j+1}}\int_{E_{min}}^{E_{max}}
\frac{d\phi}{dE}\frac{d^2\tilde{\sigma}^{B}(x,y,E)}{dxdy} dxdydE
}\,.
\label{bebc-uncorr}
\end{equation}
Then we inserted $N_{ijk}^{uncorr}$ into eq.~(\ref{encoreune-bis})
and constructed the `experimental' Born cross section
\begin{equation}
  \frac{d^2\sigma^{B}_{exp}(x^c_i,y^c_j,E^c_k)}{dxdy}= 
\frac{N_{ijk}^{uncorr}}{C} \cdot 
  \frac{
  d\tilde{\sigma}^{B}(x^c_i,y^c_j,E^c_k)/dxdy}{
  \int_{x_i}^{x_{i+1}}\int_{y_j}^{y_{j+1}}\int_{E_{min}}^{E_{max}}
  \frac{d\phi}{dE}\frac{d\tilde{\sigma}^{B+R}(x,y,E)}{dxdy}dE\,dx\,dy}\, .
\label{cross-section-bebc}
\end{equation}

 To compute the correction factors,  
 the beam energy flux function of \cite{katz}
was used\footnote{We thank
 U. Katz for having provided us the FORTRAN code computing the 
beam energy fluxes used in  \cite{katz}.} and the kinematic cuts
applied in the data analysis were taken into account:
 $Q^2>2 $GeV$^2$, $W^2>4 $GeV$^2$, $5$ $<E<160$ GeV and 
$E_\mu>5$ GeV ($E_\mu$ being the scattered muon energy). 
  
One should  notice that: {\it i)} 
The beam energy flux is not known absolutely, hence 
 the absolute normalisation of the data has to be determined by the
global fit. {\it ii)} The systematic errors are not available. 

\subsection{Nuclear effects}\label{Nuclear-targets}
 
Most of the neutrino DIS data come from experiments 
with nuclear targets (deuteron or heavy nuclei). 
Thus they have to be corrected for nuclear effects, 
which are known to be quite relevant 
(see for instance \cite{nuclear-fe,michele}).

\subsubsection{Deuteron}
 
We correct the calculated nucleon 
structure functions 
for nuclear binding, Fermi motion and off-shell effects
using the results of the covariant
approach of Melnitchouk et al. \cite{melnitchouk}\footnote{We 
thank W.~Melnitchouk for having
provided us the computer code of his calculation.}. 
The neutrino structure functions 
$F_1^{\nu D}$, $F_2^{\nu D}$ and $F_3^{\nu D}$ are  
treated analogously. 

The calculation of \cite{melnitchouk} 
describes only 
the  high-$x$ ($x\gtrsim 0.2$)
 behaviour of the deuteron structure functions. At small $x$ other 
mechanisms are at work (anti-shadowing and shadowing) but they are 
negligible in the $x$ and $Q^2$ regions of the deuteron 
data entering our analysis \cite{shadowing}.

\subsubsection{Heavy nuclei}
\label{heavy-nuclei}

All CDHSW and a large fraction of CDHS data are 
obtained from scattering off Iron nuclei. Since the 
theoretical understanding of nuclear effects in 
heavy nuclei is still uncertain and model dependent
\cite{nuclear-fe,michele}, we adopt an empirical 
procedure to perform the nuclear corrections.

 The experimental $\nu(\bar{\nu}) Fe$ differential cross sections 
are fitted to 
\begin{equation}
d\sigma^{\nu(\bar{\nu})Fe}= 
\frac{d\sigma_{iso}^{\nu(\bar{\nu})Fe}}{R_{iso}^{\nu(\bar{\nu})}}
\;, 
\label{sigma-fe}
\end{equation}
where $R_{iso}^{\nu(\bar{\nu})}$ is the correction factor for the 
non-isoscalarity of Iron
\begin{equation}\label{ison}
R_{iso}^{\nu(\bar{\nu})} =  
\frac{(d\sigma^{\nu(\bar{\nu})p}
+ d\sigma^{\nu(\bar{\nu})n})/2}{(Z \, d\sigma^{\nu(\bar{\nu})p} + 
(A-Z) \, d\sigma^{\nu(\bar{\nu})n})/A}
\;\;\;\; (A= 55.8, \; Z=26)\;, 
\end{equation}
and 
$d\sigma_{iso}^{\nu(\bar{\nu})Fe}$ is the `isoscalar' 
Iron cross section which incorporates the nuclear 
corrections. Thus 
\begin{equation*}
d\sigma_{iso}^{\nu(\bar{\nu})Fe}= 
d\sigma^{\nu(\bar{\nu})D} 
\cdot 
R_{nucl}^{iso} 
\end{equation*}
where 
 $R_{nucl}^{iso}$ is a function embodying the 
nuclear effects on an isoscalar target.  
$R_{nucl}^{iso}$ consists of two 
factors 
\begin{equation}
R_{nucl}^{iso} = R_{Fe/D} \cdot R_{iso}^{\ell}
\;.  
\label{nucl}
\end{equation}
The first factor is the $Fe/D$ structure function ratio 
\begin{equation}
R_{Fe/D} = \frac{F_2^{Fe}}{F_2^{D}} \;, 
\label{FeDratio}
\end{equation}
which is obtained 
from a fit to the published 
experimental data on 
$F_2^{Fe}/F_2^{D}$, uncorrected for isoscalarity.  
The second factor contains the isoscalarity corrections
\begin{equation} 
R_{iso}^{\ell} = \frac{(F_2^{\ell p} + 
F_2^{\ell n})/2}{(Z \, F_2^{\ell p} + (A-Z) \, F_2^{\ell n})/A} \;. 
\label{isoscalar}
\end{equation}

$R_{Fe/D}$
 is a function of $x$ only because there is 
no 
experimental evidence of a significant $Q^2$ dependence
(for a recent study, see \cite{nmc-iron}).  
Theoretically, a higher-twist ({\it i.e.} power-like) 
$Q^2$ dependence is expected at small $x$ 
and $Q^2 \sim 1$ GeV$^2$ 
\cite{kwie,melnitchouk-shadowing}, but at larger $Q^2$, 
 in the region covered by our analysis, shadowing is 
a scaling phenomenon \cite{kolya,io-shadowing}.  

The small-$x$ ($x <0.1$) 
 $\nu Fe$ and  $\bar{\nu} Fe$ are excluded in our analysis.  
The reason is that in this region 
there are a number of uncertainty sources affecting the 
 determination of $R_{nucl}$.  
First of all, there is an unsolved 
discrepancy between the two measurements $R_{Fe/D}$ at small-$x$,
 namely between
E665 \cite{e665-iron} and NMC  \cite{nmc-iron}. Second,  
 the use of the {\em charged-lepton} DIS data to determine the 
${Fe/D}$ ratio of {\em neutrino} cross sections in 
eq.~(\ref{sigma-fe}) 
is justified by the BEBC $Ne$ target results \cite{bebc-shadowing}
being in good agreement with the NMC results on C \cite{nmc-iron} 
(see \cite{exp}), but the 
 situation is experimentally not so clear  for $x\lesssim 0.1$,  
where different nuclear corrections for charged lepton and 
neutrino structure functions are expected
from a theoretical point of view \cite{marage}. 
Finally, the $F_2$ and $xF_3$ corrections 
 might be different at small $x$ \cite{kumano}. Thus
the cut at $x=0.1$ removes from our fits the more controversial 
kinematic region as for  nuclear corrections.  

\begin{figure}[htbp]
  \begin{center}
    \leavevmode
\includegraphics*[width=6in]{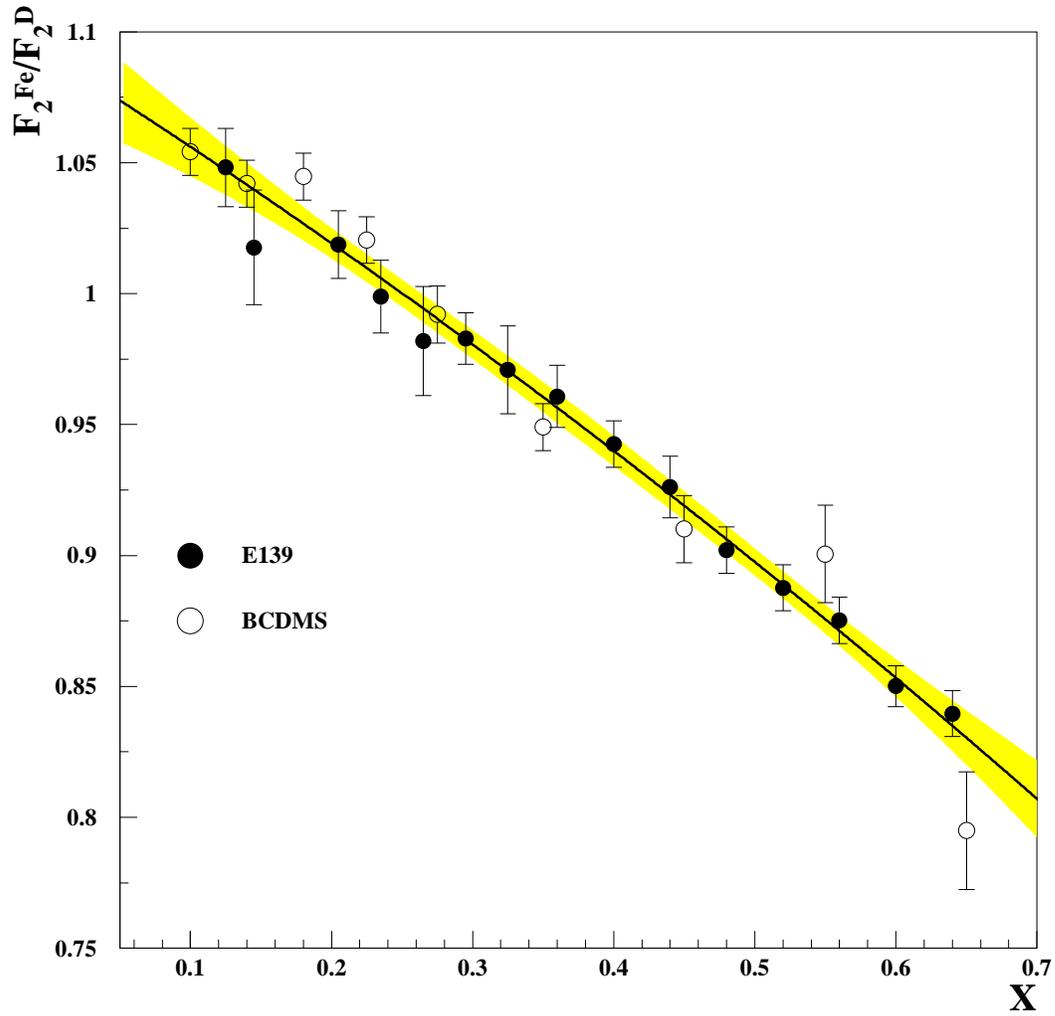}
    \caption{ $Fe/D$ structure function ratio. The full line is the result
of a second order polynomial fit and 
 the shaded area is the corresponding one-standard-deviation error band
 (see section 
\ref{Nuclear-targets}).
The  error bars correspond to the quadratic sum of statistical
 and systematic errors.}
    \label{fig-iron/D}
  \end{center}
\end{figure}

We have then performed a fit to the
 SLAC \cite{slac-iron-E139} BCDMS \cite{bcdms-iron}
 $F_2^{\mu Fe}/F_2^{\mu D}$ data in the range 
$ 0.1 < x < 0.65$. 
For the function $R_{Fe/D}(x)$ we chose the empirical form
\begin{equation}
R_{Fe/D}(x)=\alpha_1+\alpha_2 x
 + \alpha_3 x^2
\label{FeD}
\end{equation}
where $\alpha_i$, $i=1,2,3$ are the free parameters of the fit.
 The parameters
 $\bar{\alpha}_i$ minimising the $\chi^2$ of this fit (we shall 
denote it 
$\chi^2_{Fe/D}$) are given in Table~4. 
 The result of the fit is shown in fig.~\ref{fig-iron/D} together with
the one-standard-deviation error band.

 In fig.~\ref{isocor}.a,b we plotted the isoscalarity and 
nuclear+isoscalarity corrections computed from the structure functions
 obtained with our fit (see section \ref{results}).  

\begin{table}[htbp]
  \begin{center}
    \leavevmode
\begin{small}
    \begin{tabular}{|c|c|c|}
\hline
 \small{$\bar{\alpha}_1$}&\small{$\bar{\alpha}_2$} &
  \small{$\bar{\alpha}_3$} \\ 
 \small{1.091  $\pm$ 0.021}&\small{-0.34 $\pm$ 0.12}
&\small{-0.09 $\pm$ 0.16}\\
\hline
\end{tabular}
\end{small}
    \label{tab:param-iron}
\caption{\small Results of the minimisation of 
$\chi^2_{Fe/D}$ (see text). 
}
  \end{center}
\end{table}

\section{The analysis}\label{qcd-analysis}

\subsection{Data entering the fit}\label{data}

Our analysis includes the neutrino and anti-neutrino 
cross section data of BEBC \cite{bebc}, CDHS \cite{cdhs}
and CDHSW \cite{cdhsw}. 
The number of collected events, the number of points 
and the kinematic domain covered by these experiments  
are listed in Table~\ref{table-neutrino-data}. 

Besides the neutrino data, discussed at length
in 
section~\ref{Re-evaluation}, 
the bulk of data entering our fits consist of 
structure functions from various charged-lepton DIS
experiments: BCDMS \cite{bcdms}, H1 \cite{h1-94}, NMC \cite{nmc}
(see Table~\ref{table-lepton-data}). 

As for BCDMS, we do not use the merged 
structure function data, obtained by putting together 
measurements at different beam energies, but rather 
the data on the reduced cross section
\begin{equation}
\frac{Q^2}{4 \pi \alpha_{em}^2 M_N E}\, 
\frac{1}{Y}\, \frac{d^2\sigma}{dxdy}\,,\;\;\;\;\;
 Y=1+(1-y)^2-\frac{M_Nxy}{E}\, ,
\end{equation}
for each beam energy (we retrieved the original data from
 ref. \cite{bcdms_cern}). 

\begin{table}[htbp]
  \begin{center}
    \leavevmode
    \begin{tabular}{|c|c|c|c|c|}
\hline
 & \small{events} &  \small{$\Delta x$}&
 \small{$\Delta Q^2$ (GeV$^2$)}& \small{$\Delta y$} \\\hline
\small{CDHSW($\nu Fe$)} & \small{640\,000} &  
\small{$0.015-0.65$}&\small{$0.2-240$} &\small{$0.037-0.87$} \\
\small{CDHSW($\bar{\nu} Fe$)}&\small{550\,000} &" &" &" \\
\small{CDHS($\nu H$)} & \small{2\,100} & "&" &" \\
\small{CDHS($\bar{\nu} H$)}&\small{1\,100} &" &" &" \\
\small{BEBC($\nu D$)} & \small{12\,100 } &  
\small{$0.03-0.65$}&\small{$3-75$} &\small{$0.05-0.95$} \\
\small{BEBC($\bar{\nu} D$)}&\small{5\,400 }& "&" &" \\
\small{BEBC($\nu H$)} & \small{9\,800 } &
\small{$0.03-0.65$}&\small{$3-75$} &\small{$0.05-0.95$} \\
\small{BEBC($\bar{\nu} H$)}&\small{4\,900 }& "&" &" \\\hline
    \end{tabular}
    \caption{\small Number of events and limits of the kinematic domain
covered by the neutrino experiments considered  in our analysis.}
    \label{table-neutrino-data}
  \end{center}
\end{table}

\begin{table}[htbp]
  \begin{center}
    \leavevmode
    \begin{tabular}{|c|c|c|}
\hline
 &   \small{$\Delta x$}&
 \small{$\Delta Q^2$ (GeV$^2$)} \\ 
\hline
\small{NMC ($\mu p$, $\mu D$)} & \small{$0.008-0.5$} &  
\small{$0.8-60$}  \\
\small{BCDMS ($\mu p$)} & \small{$0.07-0.75$} &
\small{$7.5 - 230$} \\
\small{H1-94 ($ep$)}& \small{$3.2 \cdot 10^{-5}-0.32$} & 
                           \small{$1.5-5000$} \\
\small{E605 (DY)} & \small{$0.12-0.4$} &  
\small{$22.6-248$}  \\
\small{NA51 (DY)} & \small{$0.18$} &  
\small{$27.2$}  \\
\small{E866 (DY)} & \small{$0.036-0.312$} &  
\small{$30-164$}  \\
\hline
    \end{tabular}
    \caption{\small Kinematic range of the charged-lepton DIS 
and DY data sets entering 
our fits.} 
    \label{table-lepton-data}
  \end{center}
\end{table}

To avoid a redundancy of data sets 
and, most of all, to limit
 the number of experimental parameters in the 
$\chi^2$ minimisation, the CDHS Iron data,
which are in agreement with the CDHSW data, 
do not enter the fit. For the same 
reasons, the ZEUS $F_2$ data \cite{zeus-f2} are not included 
(ZEUS and H1 1994
 data are consistent within $1-2\%$,  as shown by a recent pQCD analysis
\cite{franck}).

 Drell-Yan (DY) data are introduced in order to constrain 
the non-strange sea.
 Three measurements are used. 
We fitted  
the differential
cross section  
for 
the reaction $p \, Cu\rightarrow \mu^+\mu^- 
X$ measured by 
E605 \cite{e605}.  
The cross section is calculated at NLO\footnote{ We
 thank W.~van Neerven for
having provided us the code computing the order $\alpha_s$ Wilson
 coefficients published in \cite{neerven-dy}.}. 
 For Drell-Yan processes  the charm contribution is small
 \cite{neerven-dy-charm} and is  neglected in our analysis. No higher 
 twist corrections are required \cite{webber} since the kinematic domain
 covered by E605 avoids the phase space boundaries 
where these 
corrections  are expected to be important\footnote{This
 is not the case for the latest 
E772 Drell-Yan data 
 \cite{e772} and that is why we do not consider them here.}.
 
The other Drell-Yan results that we use are the 
measurements of  
 the DY asymmetry in $pp$ and $pD$ collisions
 by NA51 \cite{na51} and by E866
\cite{e866}. These data constrain the 
ratio $\bar{u}/\bar{d}$.

The kinematic cuts applied to the DIS data entering the fit are:  
\begin{itemize}

\item
$Q^2\ge 3.5$ GeV$^2$ and $W^2\ge 10$ GeV$^2$. 
\end{itemize}
 In this region higher-twist  effects are 
negligible \cite{virchaux-alphas}. 
Target mass corrections \cite{tmc} are also very small 
  but have been  taken into account in our calculations. 

Due to the $W^2$ and $Q^2$ cuts 
 neither 
the E665 \cite{e665-f2}, nor the H1 (1995) \cite{h1-1995}, nor
 the SLAC  \cite{slac},
 nor the ZEUS (1995) $F_2$ data \cite{zeus-1995} enter our fits.

Finally, we  reject 
 the CDHSW
 data with $x<0.1$. The reason for this  cut is threefold: 
 {\it i)} the systematic errors  in the low-$x$
region are large \cite{vallage}; {\it ii)}
 the nuclear corrections at small $x$ are not completely 
under control, as discussed in section~\ref{Nuclear-targets}; 
{\it iii)} at low-$x$ the CDHSW 
results disagree with 
the CCFR findings for the cross sections \cite{yang}
and for the structure functions \cite{seligman}.

\subsection{Fitting procedure}\label{procedure}

The main steps of our fitting procedure are summarised below.
For each iteration:

\begin{enumerate}
\item 
 The pdf's are parametrised at a given value of the momentum 
transfer denoted $Q_0$. 
We choose $Q_0^2= 4 $ GeV$^2$.  
\item
 The DGLAP equations are solved numerically in the $x$-space 
\cite{nous-method} (see \cite{nous-hera} for a comparison
of different NLO evolution codes).
\item
 The evolved pdf's are convoluted with the Wilson
 coefficients
to obtain 
the structure functions 
 (see section~\ref{QCD-phenomenology}). 
\item
 Assuming that all experimental
uncertainties are normally distributed the $\chi^2$ is 
computed
\begin{eqnarray}\label{lechi2}
\chi^2
&=&\sum_{exp} \sum_{dat}
 \frac{[{\cal O}^{dat}_{exp}-
{\cal O}^{ fit}\times(1-\nu_{exp}\sigma_{exp}
-\sum_k\delta^{dat}_k(s^{exp}_k))
]^2}
{\sigma_{dat,stat}^2+\sigma_{dat,uncor}^2}\nonumber\\
&+&\sum_{exp}\nu_{exp}^2  
+\sum_{exp}\sum_k (s^{exp}_k)^2 \nonumber \\
&+& \sum_{i,j=1}^2 
[(a^\nu_i-\bar{a}^\nu_i)M_{ij}^{\nu} (a^\nu_j-\bar{a}^\nu_j)  
+(a^{\bar{\nu}}_i-\bar{a}^{\bar{\nu}}_i)M_{ij}^{\bar{\nu}}
 (a^{\bar{\nu}}_j-\bar{a}^{\bar{\nu}}_j)], 
\end{eqnarray}
where ${\cal O}$ stands for the observables  
(structure functions and differential cross sections). 
The first two sums run over the data ($dat$) of the
various experiments ($exp$); 
 $\sigma_{exp}$ is the relative overall normalisation uncertainty; 
 $\sigma_{dat,stat}$ and $\sigma_{dat,uncor}$ are the statistical error 
 and the uncorrelated systematic error, 
respectively,  corresponding to the datum $dat$;
 $\nu_{exp}$ is the number of standard deviations corresponding to
 the overall normalisation of the experimental sample $exp$;
 $\delta^{dat}_k(s^{exp}_k)$ is the relative shift of the datum $dat$
 induced by a change by $s^{exp}_k$ standard deviations of
 the $k^{th}$ correlated systematic uncertainty source
  of the experiment $exp$. It is estimated by 
\begin{eqnarray}
\delta^{dat}_k(s^{exp}_k)=
\frac{{\cal O}^{dat}_{exp}(s^{exp}_k=+1)-{\cal O}^{dat}_{exp}(s^{exp}_k=-1)}
{2{\cal O}^{dat}_{exp}}\, s^{exp}_k+\nonumber\\
\biggl[
\frac{{\cal O}^{dat}_{exp}(s^{exp}_k=+1)+{\cal O}^{dat}_{exp}(s^{exp}_k=-1)}
{2{\cal O}^{dat}_{exp}}-1
\biggr] \,(s^{exp}_k)^2\nonumber
\end{eqnarray}
where ${\cal O}^{dat}_{exp}(s^{exp}_k=\pm1)$ is the experimental determination
of ${\cal O}^{dat}_{exp}$ obtained varying by $\pm 1 \sigma$ the $k^{th}$
source of uncertainty.
 The last term in eq. (\ref{lechi2}) has already been discussed in
section~\ref{CDHSW-measurements} (eq.~(\ref{anuanubar})). 

  Notice that even though the parameters $\nu_{exp}$, $s^{exp}_k$,
  $a^\nu_i$ and $a^{\bar{\nu}}_i$ are 
obtained from  the global 
 $\chi^2$ minimisation, 
 they do not enter in the counting  of the degrees of freedom
 since they are determined by the counter-terms.

The correlated systematic 
 uncertainties 
are
taken into account whenever the information  about them is 
available. This is the case for 
H1 \cite{h1-94}, BCDMS \cite{bcdms_cern} and NMC \cite{nmc}.
 For CDHS and BEBC no information is available and  
 for E605 \cite{e605} 
 the uncorrelated systematic uncertainties are dominating.
 The systematic uncertainties of this later experiment  
has then been added in quadrature
 and included in  $\sigma^2_{dat,uncor}$ of eq.~(\ref{lechi2}).
 As  already pointed out in section~\ref{BEBC-measurements}, the 
 systematic uncertainties of the BEBC data are not known.
 Since these data cover the same region as the 
 CDHS data, for both data sets 
we have taken into account only the statistical
 uncertainties (this has a 
negligible effect on the $\chi^2$ minimisation, since the statistical 
significance of these samples is rather small).

\item

 The MIGRAD algorithm of the MINUIT program \cite{minuit} 
is used to minimise
 the $\chi^2$. 

\end{enumerate}

Given the 
importance of nuclear effects in the treatment of the neutrino data,
in parallel to the main fit, we performed, as a check, another fit, 
in which the nuclear parameters 
$\alpha_i$ of eq.~(\ref{FeD}) are not constrained to the values 
$\bar \alpha_i$ obtained by the independent parametrisation of the
$Fe/D$ structure function ratio described in section~\ref{heavy-nuclei}, 
but are readjusted a posteriori. 
This is done
by adding the counter-term 
\begin{equation}\label{chi2fe}
\sum_{i,j=1}^3(\alpha_i-\bar{\alpha}_i)M_{ij}^{Fe/D}
(\alpha_j-\bar{\alpha}_j)
\end{equation}
to the $\chi^2$ expression of the global pQCD analysis.     
 In eq.~(\ref{chi2fe})
$M_{ij}^{Fe/D}=(1/2)\, \partial^2\chi^2_{Fe/D}/\partial 
\alpha_i\partial\alpha_j$. 
We found that the two fits give very similar 
results.

\subsection{The parametrisation}\label{fit-procedure}

Imposing the isospin symmetry leads to 
the following relations among the pdf's: 
 $u^p=d^n\equiv u$, $ d^p=u^n\equiv d$,  
 $\bar{u}^p=\bar{d}^n\equiv\bar{u}$,
 $\bar{d}^p=\bar{u}^n\equiv\bar{u}$, 
$s^p=s^n\equiv s$, $\bar s^p = \bar s^n \equiv \bar s$.

In our main fit, that we call {\tt fit1},  
the pdf's $u_v \equiv u - \bar u, 
d_v \equiv d - \bar d, \bar u, \bar d, s, \bar s$ and 
$g$ (the gluon density) are parametrised at 
$Q_0^2 = 4$ GeV$^2$ as follows: 
\begin{eqnarray}
xu_v(x,Q_0^2) = 
A_{u_v} \, x^{B_{u_v}} \, 
(1-x)^{C_{u_v}}\, (1+D_{u_v} \, x^{E_{u_v}}),
\label{pdf_uv} \\
xd_v(x,Q_0^2) = 
A_{d_v} \, x^{B_{d_v}} \, 
(1-x)^{C_{d_v}}\, (1+D_{d_v} \, x^{E_{d_v}}),
\label{pdf_dv} \\
x(\bar u + \bar d)(x,Q_0^2) = 
A_{+} \, x^{B_{+}} \, 
(1-x)^{C_{+}}\, (1+D_{+} \, x^{E_{+}}),
\label{pdf_+} \\
x(\bar d - \bar u)(x,Q_0^2) = 
A_{-} \, x^{B_{-}} \, 
(1-x)^{C_{-}}\, (1+D_{-} \, x),
\label{pdf_-} \\
x s(x, Q_0^2) = x \bar s(x,Q_0^2) = 
A_{s} \, x^{B_{s}} \, 
(1-x)^{C_{s}}\, (1+D_{s} \, x^{E_{s}}),
\label{pdf_s} \\
xg(x,Q_0^2) = 
A_{g} \, x^{B_{g}} \, 
(1-x)^{C_{g}}\, (1+D_{g} \, x^{E_{g}}).
\label{pdf_g} 
\end{eqnarray}
This parametrisation form is similar to that used
 in \cite{cteq-input} (we refer to this article for a justification 
of this choice). 

Generally it is also assumed $s = \bar s$. 
The data samples used in 
 the existing global analyses \cite{cteq,mrs,grv}  
 cannot resolve in fact $s$ and $\bar{s}$ independently. 
In our case, the
information coming from 
 neutrino and anti-neutrino differential cross sections 
allows testing the hypothesis $s=\bar{s}$.
We thus performed another fit, called 
{\tt fit2}, allowing 
for a charge asymmetry 
in the strange sea, $s \ne \bar s$.

Some of the  parameters in 
eqs.~(\ref{pdf_uv}-\ref{pdf_g}) are determined by physical constraints.
 One normalisation factor, say $A_g$, is fixed by the momentum sum rule, 
 $\int_0^1 (xg+x \sum_i (q_i + \bar q_i)) \, dx=1$. 
The two normalisation parameters
 $A_{u_v}$ and $A_{d_v}$ are fixed by 
 the number sum  rules $\int_0^1 u_v dx=2$ and $\int_0^1 d_v dx=1$.  

While the intermediate-$x$ and large-$x$ shape 
of the strange distribution is well constrained 
by the data entering the fit, the small-$x$ behaviour is not. 
 Thus we set $B_s=B_{+}$.  We also set $B_{u_v} = 
B_{d_v} = B_-$, as suggested by Regge theory. It is to be mentioned that, from
 a statistical point of view, 
 these constraints do not worsen the $\chi^2$ and are also required in order
to get an invertible second derivative $\chi^2$ matrix.

\section{Results}\label{results}

We present now the results of our fits. 
The strong coupling is fixed at the value $\alpha_s(M_Z^2) =0.120$
which is close to the world average \cite{PDG}.

\begin{table}[htbp]
  \begin{center}
\begin{small}
    \begin{tabular}{|l|l|l|l|c|}
\hline
Experiments&$\#$ pts &  $\chi^2$  & $\chi^2$  &$\#$ exp  \\
     &   & {\tt fit1}  & {\tt fit2}   & param \\ \hline\hline
BCDMS(100) $\sigma^{\mu p}$&94& 108.0& 110.8&12\\
BCDMS(120) $\sigma^{\mu p}$&99& 81.6& 81.0&\\
BCDMS(200) $\sigma^{\mu p}$&79& 92.7& 90.7&\\
BCDMS(280) $\sigma^{\mu p}$&76& 89.7& 87.6&\\ 
BCDMS(120) $\sigma^{\mu D}$&99& 96.9& 93.1&\\ 
BCDMS(200) $\sigma^{\mu D}$&79& 93.5& 88.8&\\ 
BCDMS(280) $\sigma^{\mu D}$&76& 64.6& 63.7&\\\hline
BEBC $\sigma^{\nu p}$&68&  65.2&  67.6&  0\\ 
BEBC $\sigma^{\bar{\nu} p}$&49&  76.4&  76.5&  \\ 
BEBC $\sigma^{\nu D}$&70&  65.3&  65.4&  \\ 
BEBC $\sigma^{\bar{\nu} D}$&49&  49.7&  46.7& \\\hline
CDHS $\sigma^{\nu p}$&45&50.6 &  48.6&  2\\ 
CDHS $\sigma^{\bar{\nu} p}$&42&  53.7&  53.2&  \\\hline
CDHSW $\sigma^{\nu Fe}$&494& 264.9& 250.4& 5\\ 
CDHSW $\sigma^{\bar{\nu} Fe}$&492& 274.2& 277.5& \\\hline
E605(DY)&136& 104.9& 104.1&1\\\hline
E866 $A_{DY}$&11&  8.2&  8.2& 1\\ 
H1(94-svx) $F_2^{ep}$&24&  26.7 &  26.7&  7\\ 
H1(94-nvx) $F_2^{ep}$&156& 180.6& 180.3& \\\hline
NMC(90) $F_2^{\mu p}$&34&  32.6&  32.5& 16\\ 
NMC(120) $F_2^{\mu p}$&46&  67.0&  67.3& \\ 
NMC(200) $F_2^{\mu p}$&61&  99.4&  99.3& \\ 
NMC(280) $F_2^{\mu p}$&68& 106.1& 105.6& \\ 
NMC(90) $F_2^{\mu D}$&34&24.0  &  23.8& \\ 
NMC(120) $F_2^{\mu D}$&46&  50.6&  50.8&  \\ 
NMC(200) $F_2^{\mu D}$&61&  62.6&  62.7& \\ 
NMC(280) $F_2^{\mu D}$&68&  94.8&  95.6& \\\hline
NA51 $A_{DY}$&1&   2.7&   2.7&  0\\\hline\hline
Total $\chi^2$ & 2657 & 2430.8& 2405.0& 44\\ \hline
    \end{tabular}
\end{small}
\caption{\small Contribution to the 
global $\chi^2$, and number of points,
 of  the data samples entering the fits. The values of the individual 
 $\chi^2$'s do not include the normalisation and 
correlated systematic shifts. 
The last column indicates the number of experimental parameters:
 overall normalisation of the data sets and correlated systematics.
 The contributions of the 
 experimental parameters to the $\chi^2$ amount to 43.6 
 for {\tt fit1} and 40.5 for {\tt fit2}.}
    \label{tab:data-in-the-fit}
  \end{center}
\end{table}

 The contributions of the different data sample to the $\chi^2$
 are given in table~\ref{tab:data-in-the-fit}.  
The total $\chi^2$ per degree of freedom is excellent for both fits.
The amount of systematic corrections for each datum,
 that is the value of the term
 $[\nu_{exp}\sigma_{exp}+\sum_k\delta^{dat}_k(s^{exp}_k)]$
 in eq. (\ref{lechi2}), is
 between -6$\%$ and +12$\%$ for all fits.

\begin{table}[htbp]
  \begin{center}
    \leavevmode
\begin{small}
    \begin{tabular}{|l|l|l|}
\hline
 & {\tt fit1} & {\tt fit2}  \\ \hline\hline
$A_g$   &   7.72 &  7.3 \\
$B_g$   &  0.089 &    0.081\\
$C_g$   &  21.26 &   20.11\\
$D_g$   &12072&  6990\\
$E_g$   &    4.17&    4.0\\ \hline
$A_{u_v}$&     1.43 &    2.05  \\
$B_{u_v}$&    0.49&    0.55\\
$C_{u_v}$&    3.60& 3.75   \\
$D_{u_v}$&   4.47&   3.59 \\
$E_{u_v}$&    0.81&    0.97\\ \hline
$A_{d_v}$&     1.02 &     1.40 \\
$B_{d_v}$&     0.49 &    0.55 \\
$C_{d_v}$&    6.03&    6.63\\
$D_{d_v}$&   23.06&   32.63\\
$E_{d_v}$&    1.76&    2.07\\ \hline
$A_+$&    0.071&    0.075\\
$B_+$&    -0.245&    -0.240\\
$C_+$&    8.31&  8.62\\
$D_+$&    11.29&    13.30\\
$E_+$&    0.88&    0.97\\ \hline
$A_-$&    0.11&    0.12\\
$B_-$&    0.49 &    0.55  \\
$C_-$&    16.08&    16.31\\
$D_-$&   -55.62&   -58.46\\
$E_-$&     1 &    1   \\ \hline
$A_{s}$ &   0.064&    0.066\\
$B_{s}$ &     -0.245 &  -0.240  \\
$C_{s}$ &    5.31&    5.59\\
$D_{s}$ &    443&    11354\\
$E_{s}$ &    8.26&    12.04\\\hline
$A_{\bar s}$ &     0.64 &     0.066 \\
$B_{\bar s}$ &    -0.245 &   -0.240 \\
$C_{\bar s}$ &    5.31 &    5.44\\
$D_{\bar s}$ &     443 &    339\\
$E_{\bar s}$ &    8.26 &    7.39\\\hline
    \end{tabular}
\end{small}
\caption{\small Values of the parameters of the pdf's at 
$Q_0^2=4$ GeV$^2$. }
    \label{tab:result-of-fit1}
  \end{center}
\end{table}

\begin{table}[htbp]
  \begin{center}
    \leavevmode
\begin{small}
    \begin{tabular}{|c|ccccccc|}
\hline

$Q^2$& $g$ & $u_v$ & $d_v$ & $\bar u$ & $\bar d$ &$s$ &$\bar s$ \\
\hline\hline
5  &42.9$\pm$0.4&27.9$\pm$0.3&11.2$\pm$0.2&3.0$\pm$0.1&3.9$\pm$0.1&
2.1$\pm$0.2&2.1$\pm$0.2\\
GeV$^2$&42.8$\pm$0.5&27.9$\pm$0.3&11.2$\pm$0.2&2.9$\pm$0.1&3.8$\pm$0.2&
2.3$\pm$0.2&2.2$\pm$0.2
\\ \hline
20 &46.6$\pm$0.3&24.6$\pm$0.3&9.8$\pm$0.2&3.2$\pm$0.1&4.0$\pm$0.1&
2.4$\pm$0.2&2.4$\pm$0.2\\
GeV$^2$&46.5$\pm$0.4&24.7$\pm$0.2&9.9$\pm$0.2&3.1$\pm$0.1&3.9$\pm$0.1&
2.6$\pm$0.2&2.4$\pm$0.2
\\ \hline
100&49.4$\pm$0.3&22.1$\pm$0.3&8.8$\pm$0.2&3.3$\pm$0.1&4.0$\pm$0.1&
2.6$\pm$0.2&2.6$\pm$0.2\\
GeV$^2$&49.3$\pm$0.3&22.1$\pm$0.2&8.8$\pm$0.2&3.3$\pm$0.1&4.0$\pm$0.1&
2.8$\pm$0.2&2.7$\pm$0.2
  \\ \hline
  \end{tabular}
\end{small}
\caption{\small  Fraction of the 
total nucleon momentum carried by the partons for
three values of $Q^2$. The results of {\tt fit1} 
(upper row) and {\tt fit2} (lower row) are 
displayed.
 The errors are computed as explained in  Appendix II.}
    \label{momentum-fraction}
  \end{center}
\end{table}

\begin{figure}[htbp]
  \begin{center}
    \leavevmode
    \includegraphics*[width=6in]{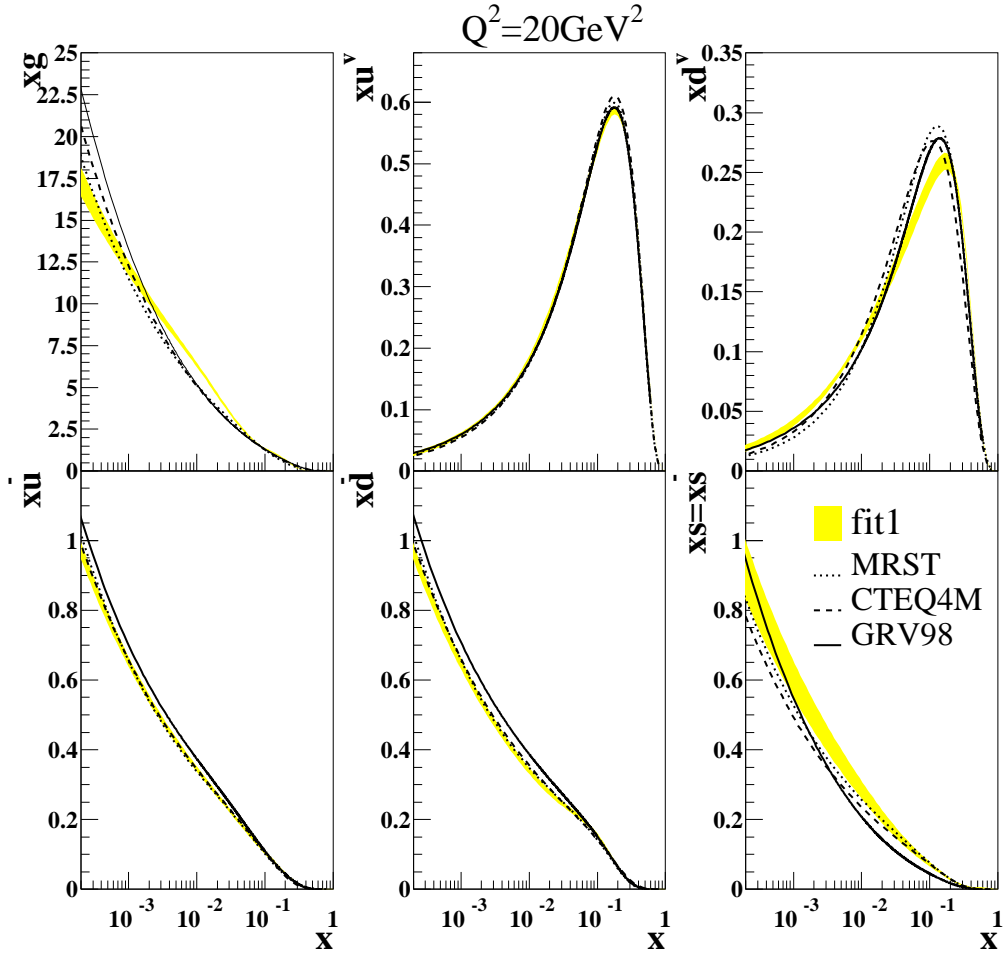}
    \caption{The parton distribution functions of 
{\tt fit1}, with their error bands, compared  
to the GRV98 (solid line), CTEQ4M (dashed line) and MRST 
(dotted line) fits. The results of {\tt fit1} 
 for  $u_v$, $\bar{u}$ and $\bar{d}$ -- hardly visible -- 
nearly coincide with the
MRST and CTEQ4M results. }
    \label{pdf99}
  \end{center}
\end{figure}

The parameters of eqs.~(\ref{pdf_uv}--\ref{pdf_g}) are listed in 
table~\ref{tab:result-of-fit1} and the parton densities  
 are shown in fig.~\ref{pdf99}, where they are compared to the 
results of the other global fits. 
The momentum 
fractions of the various partons are listed in 
table~\ref{momentum-fraction}. 

The parton distributions in fig.~\ref{pdf99} are accompanied by 
the error bands computed as explained in 
Appendix II. 
These do not take into account the 
uncertainties related to the functional choice of 
the pdf's, nor those inherent to the treatment
of the errors, which are assumed to be normally 
distributed. 
The meaning of the error bands of our pdf's is 
the following. Once a specific form for the pdf's is 
chosen and the constraints described in section~\ref{fit-procedure}
are imposed, the error bands correspond to an increase 
of the $\chi^2$ by one unit. Thus their width 
 is determined not only by the abundance and the 
precision of the data but also by the constraints on the 
pdf's. This explains why the error bands may be small even in 
kinematic regions where there are no data.  

With respect to the other parametrisations, our gluon 
density turns out to be higher at intermediate $x$. 
This can be due to the fact that we do not use the prompt photon 
data, which tend to favour a larger glue at high $x$ 
and a smaller one at intermediate $x$. These data
are still quite controversial and some of them seem to be 
in disagreement with the QCD predictions (for a recent discussion
see \cite{fontannaz}).
Also their compatibility with the DIS data 
is  still an unsettled issue. Our glue is determined by DIS  
measurements only and might be in 
slight disagreement with the prompt photon data
at medium and high $x$. We will explore this problem in a
future work where we will make use 
of an  enlarged  dataset.  

We also find a slight discrepancy in the $d_v$ distribution 
between our results and the other global fits. This is 
 not surprising, since  in our 
parametrisation we did not include the CDF
 data on the lepton asymmetry 
in $W$ production at the Tevatron \cite{cdf}. These data 
constrain the $u/d$ ratio at $x=0.05-0.1$, which is precisely 
the region where some difference can be seen between our $d_v(x)$ 
and the CTEQ and MRST distributions.

\begin{figure}[htbp]
  \begin{center}
    \leavevmode
    \includegraphics*[width=6in]{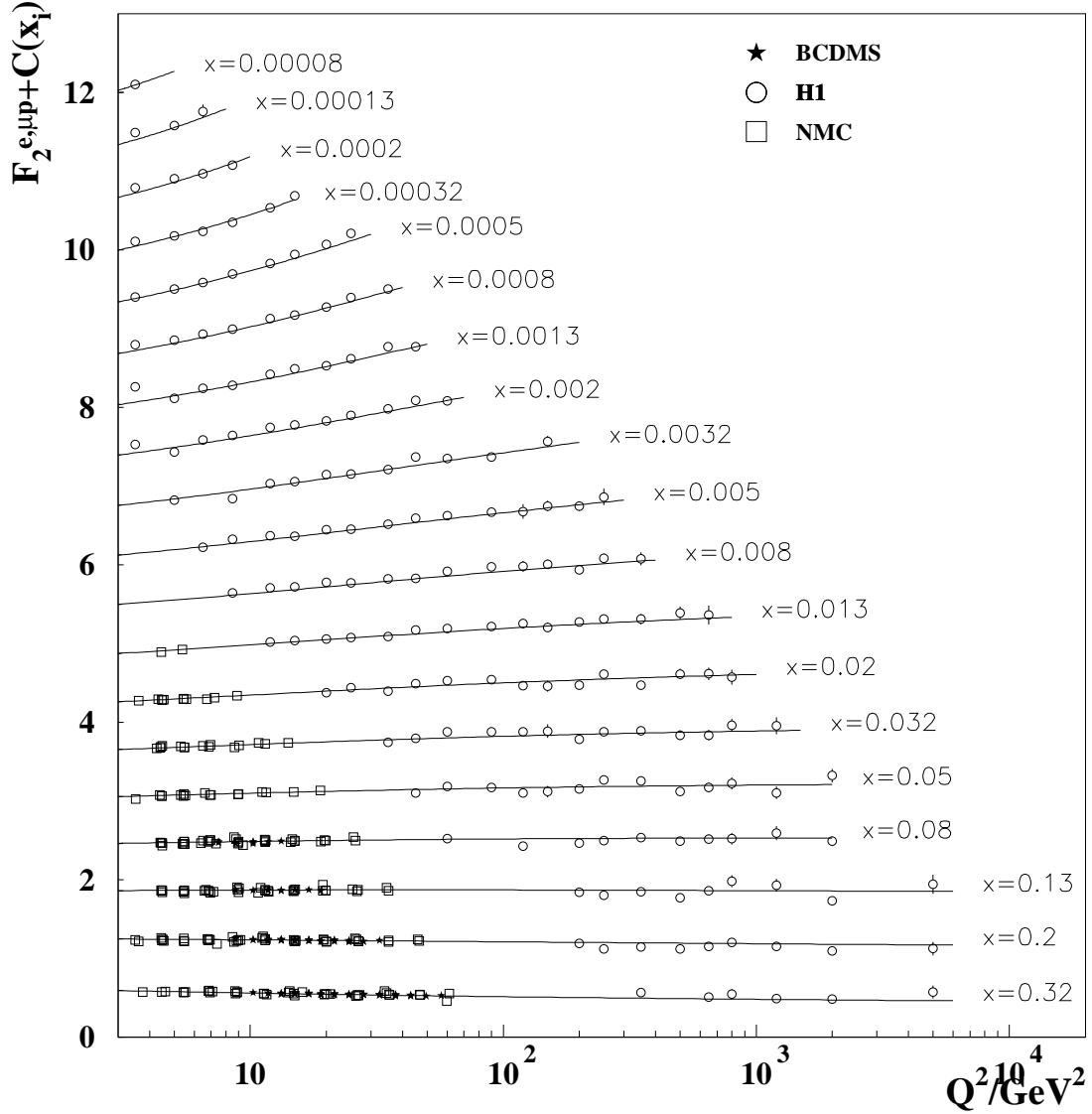}
    \caption{H1 data vs.
  {\tt fit1}. BCDMS and NMC Hydrogen target data belonging to the $x$
 domain covered by H1 are also shown.  These two data sets have
 been rebinned into the H1 $x$ bins for plotting purpose.
 The data are renormalised  by the overall
 normalisation factor determined by the fit and they are plotted 
with an
 additive bin constant $c(x_i)=0.6*(i-0.5)$
 corresponding to $x_i=\{0.32,...,8.10^{-4}\}$.
 The error bars represent the quadratic sum of
 statistical and uncorrelated systematic errors.
}
   \label{h1-data}
  \end{center}
\end{figure}

\begin{figure}[htbp]
  \begin{center}
    \leavevmode
    \includegraphics*[width=6in]{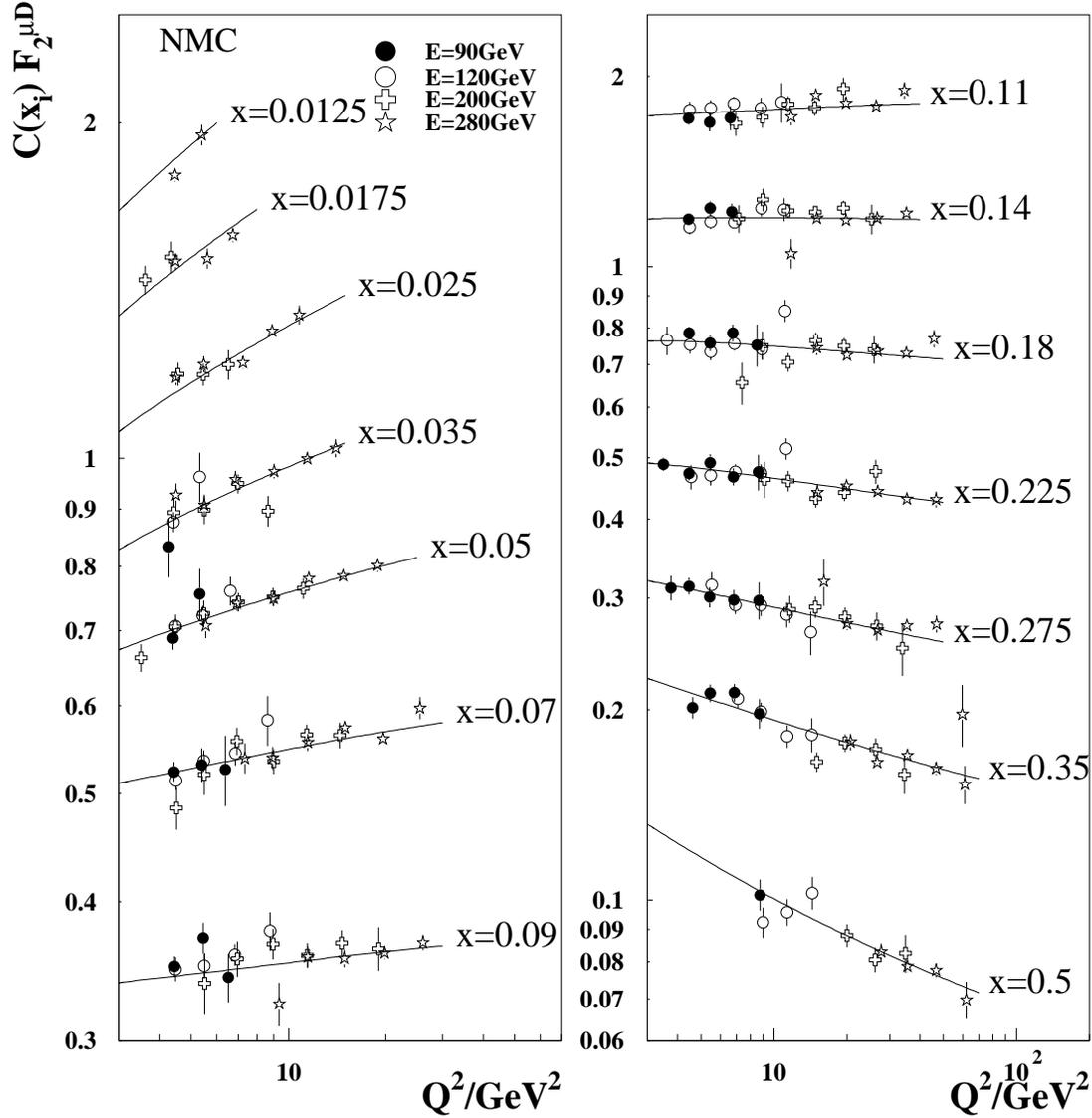}
    \caption{NMC Deuterium target data
 vs. {\tt fit1}. The different beam energy samples are shown
 separately. The data are renormalised  by the overall
 normalisation factor determined by the fit and multiplied by
 a constant $c(x_i)=\{$4.8, 4, 3.2, 2.5, 2, 1.5, 1.2, 1, 7.5, 5.2, 3.7,
2.5, 1.7, 1.2, 1, 1$\}$
 corresponding to $x_i=\{$0.0125,..., 0.07, 0.09,..., 0.5$\}$.
 The data have been re-binned into these $x_i$ bins for plotting 
purpose.
 The error bars represent the quadratic sum of
 statistical and uncorrelated systematic errors.
}
   \label{nmc-data}
  \end{center}
\end{figure}

\begin{figure}[htbp]
  \begin{center}
    \leavevmode
    \includegraphics*[width=6in]{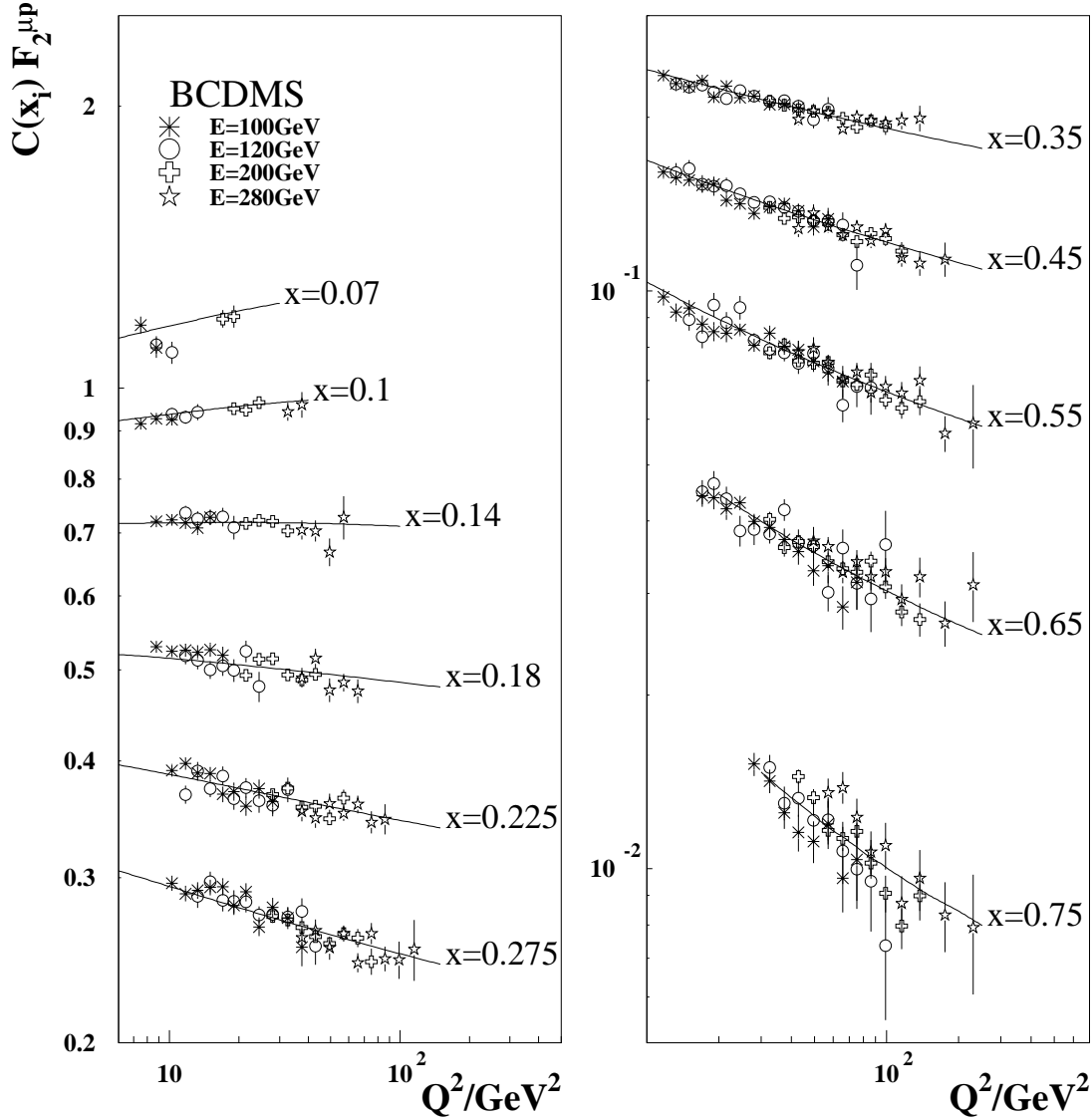}
    \caption{ BCDMS 
 hydrogen target data compared to
  {\tt fit1}. The different beam energy samples are shown
 separately.
 The data are 
renormalised by the overall normalisation factors and the 
correlated systematic
 shifts determined by the fit.
 The error bars correspond to the quadratic
 sum of statistical and uncorrelated systematic errors. As we use the BCDMS
differential cross-section data in our fits, the $F_2$ data shown in this plot
 have been determined by using $F_L$ from {\tt fit1}.
 The data are renormalised  by the overall
 normalisation factor determined by the fit and multiplied by
 a constant $c(x_i)=\{$3, 2.5, 2, 1.5, 1.2, 1, 1, 1, 1, 1, 1$\}$
 corresponding to $x_i=\{$0.07,..., 0.275, 0.35,..., 0.75$\}$.}
    \label{bcdms}
  \end{center}
\end{figure}

\begin{figure}[htbp]
  \begin{center}
    \leavevmode
    \includegraphics*[width=6in]{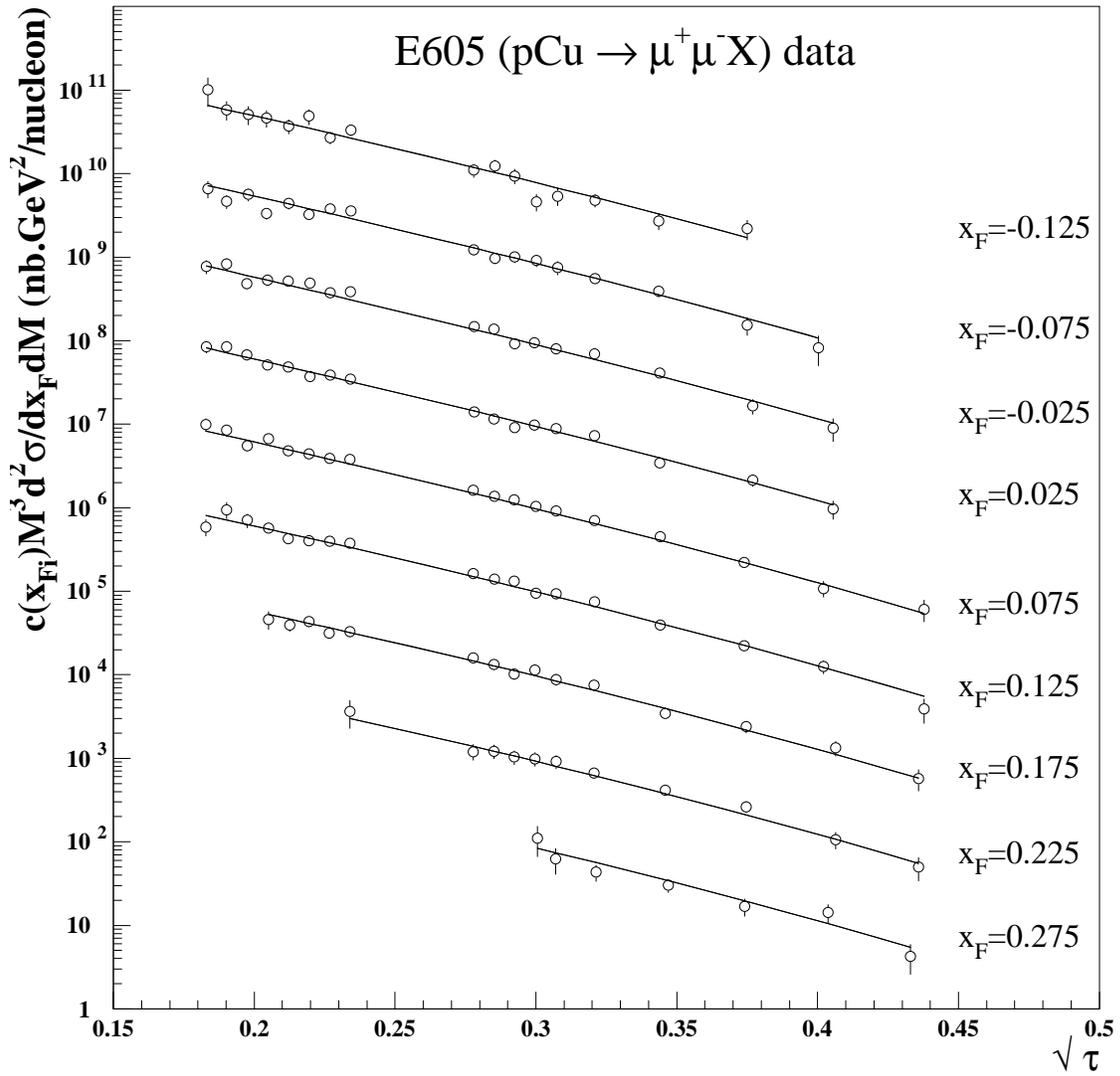}
    \caption{E605 data 
 vs. {\tt fit1}. The data have been renormalised by 
$+8\%$, as determined by {\tt fit1}. The data are multiplied 
by a constant  $c(x_{Fi})=\{10^2,...,10^{10}\}$ corresponding to
 $x_{Fi}=\{0.275,...,-0.125\}$. The error bars represent the quadratic sum of
 statistical and systematic errors.}
   \label{e605-data}
  \end{center}
\end{figure}

\begin{figure}[htbp]
  \begin{center}
    \leavevmode
    \includegraphics*[width=6in]{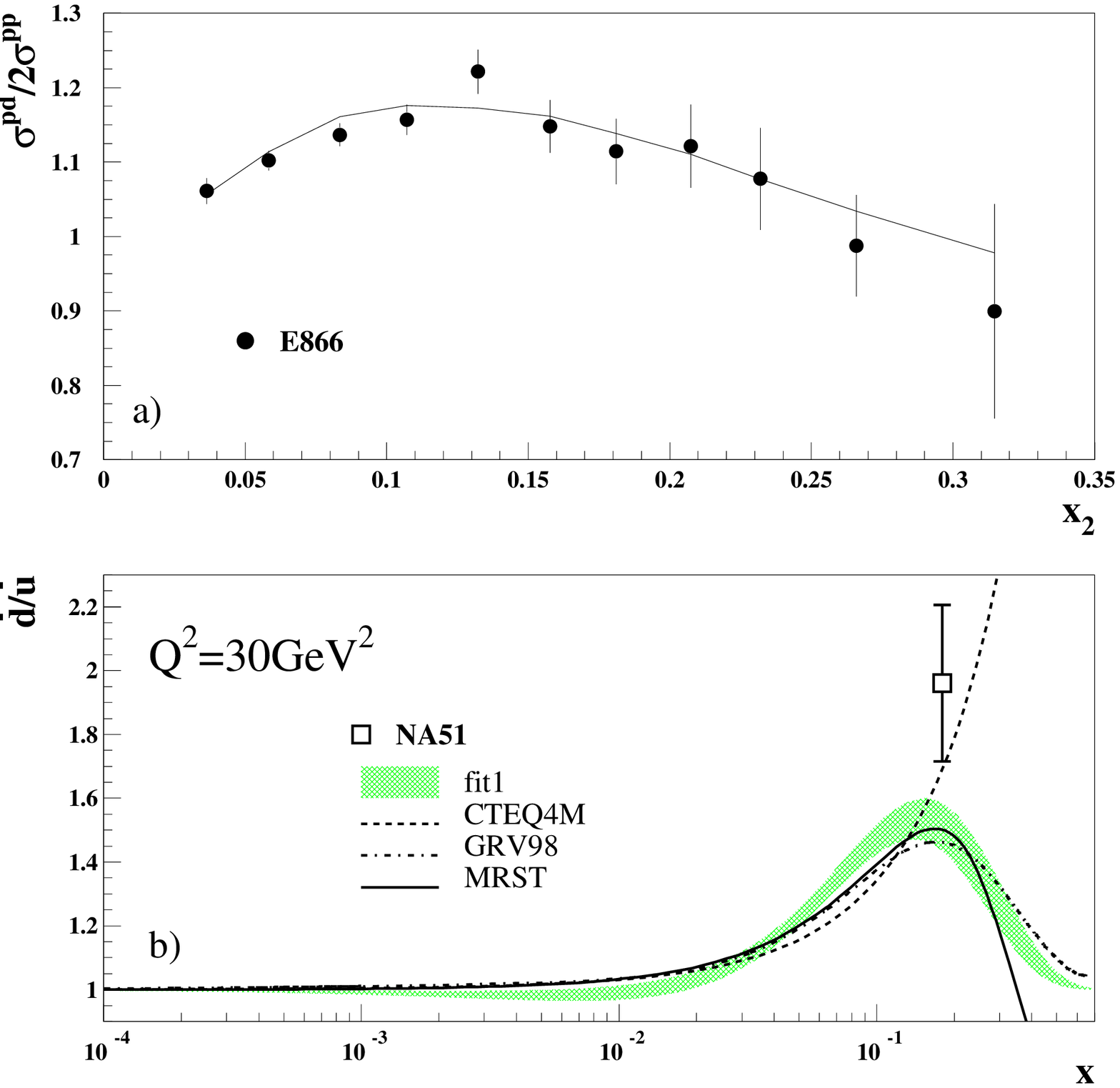}
    \caption{ a)E866 vs. {\tt fit1}.
 The data have been renormalised by $+2\%$ as determined
 in {\tt fit1}. The error bars represent the quadratic sum of
 statistical and systematic errors. Notices that the data points correspond to
 different values of $Q^2$. The line are thus an interpolation between the 
 calculations of {\tt fit1}. b) NA51's result for $\bar{u}/\bar{d}$ compared to
 {\tt fit1} and various parametrisations (CTEQ4M does not include the E866
 data).}
    \label{e866-data}
  \end{center}
\end{figure}

The structure function $F_2$ measured in different 
charged lepton DIS experiments is shown in figs.~\ref{h1-data},
 \ref{nmc-data} and \ref{bcdms} with the curves of {\tt fit1}.  
Our fits for the  Drell-Yan data are presented 
in figs.~\ref{e605-data} and
 \ref{e866-data}.
 An excellent overall agreement is observed.
In fig.~\ref{e866-data}.b we plotted the 
 ratio
$\bar{d}/\bar{u}$ at $Q^2=30$GeV$^2$,  
together with its
 error band. Notice that the fit, which is dominated by the E866 data,  
yields a $\bar{d}/\bar{u}$ ratio which lies below the 
NA51 determination.

\begin{figure}[htbp]
  \begin{center}
    \leavevmode
    \includegraphics*[width=6in]{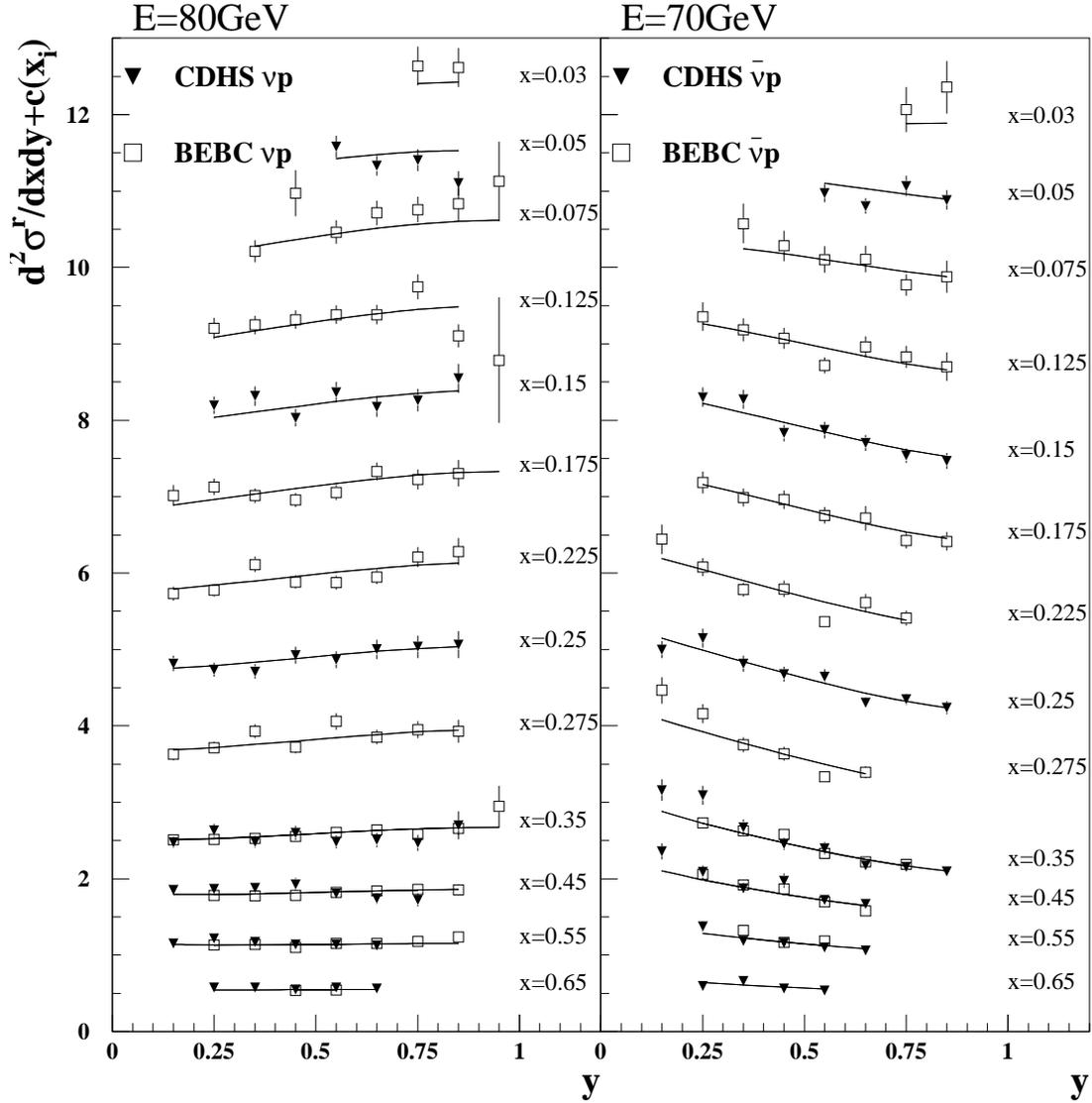}
    \caption{ CDHS and BEBC Hydrogen data vs. {\tt fit1}.
 The BEBC data have been rebinned into the CDHS bins
 for plotting purpose. The data are renormalised  by the overall
 normalisation factor determined by the fit and they are plotted 
with an additive bin constant
 $c(x_i)=\{11, 10, 9, 8, 7, 6, 5, 4, 3, 2, 1.5, 1, 0.5\}$
 corresponding to
 $x_i=\{0.03,...,0.65\}$.
 The error bars represent the statistical errors.}
    \label{fig-cdhs}
  \end{center}
\end{figure}

\begin{figure}[htbp]
  \begin{center}
    \leavevmode
    \includegraphics*[width=6in]{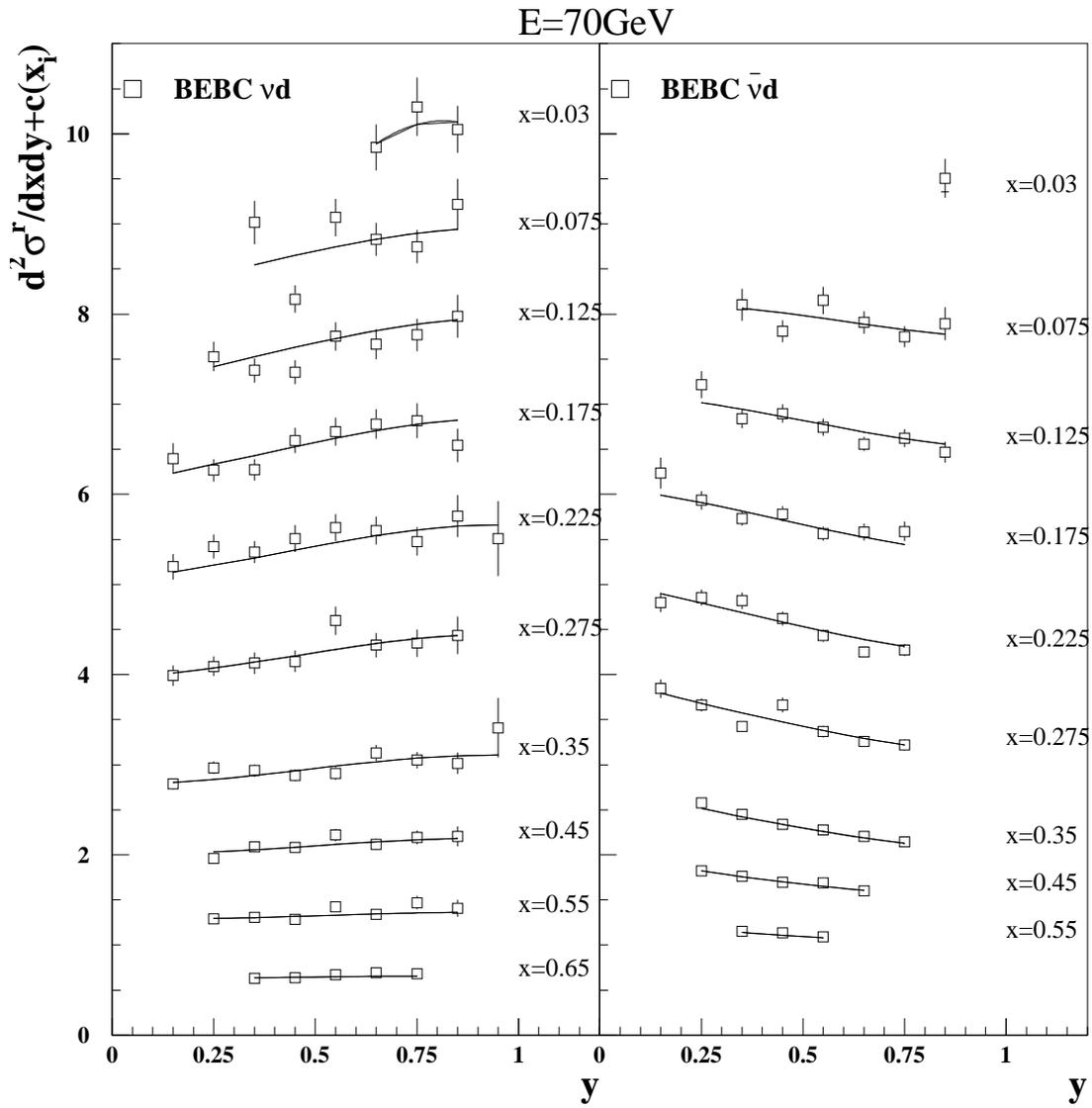}
    \caption{ Same as fig.~\ref{fig-cdhs} but for BEBC Deuterium data.
 Here the  additive bin constants are
 $c(x_i)=\{8.5, 7, 6, 5, 4, 3, 2, 1.5, 1, 0.5 \}$.}
    \label{fig-bebc}
  \end{center}
\end{figure}

\begin{figure}[htbp]
  \begin{center}
    \leavevmode
    \includegraphics*[width=6in]{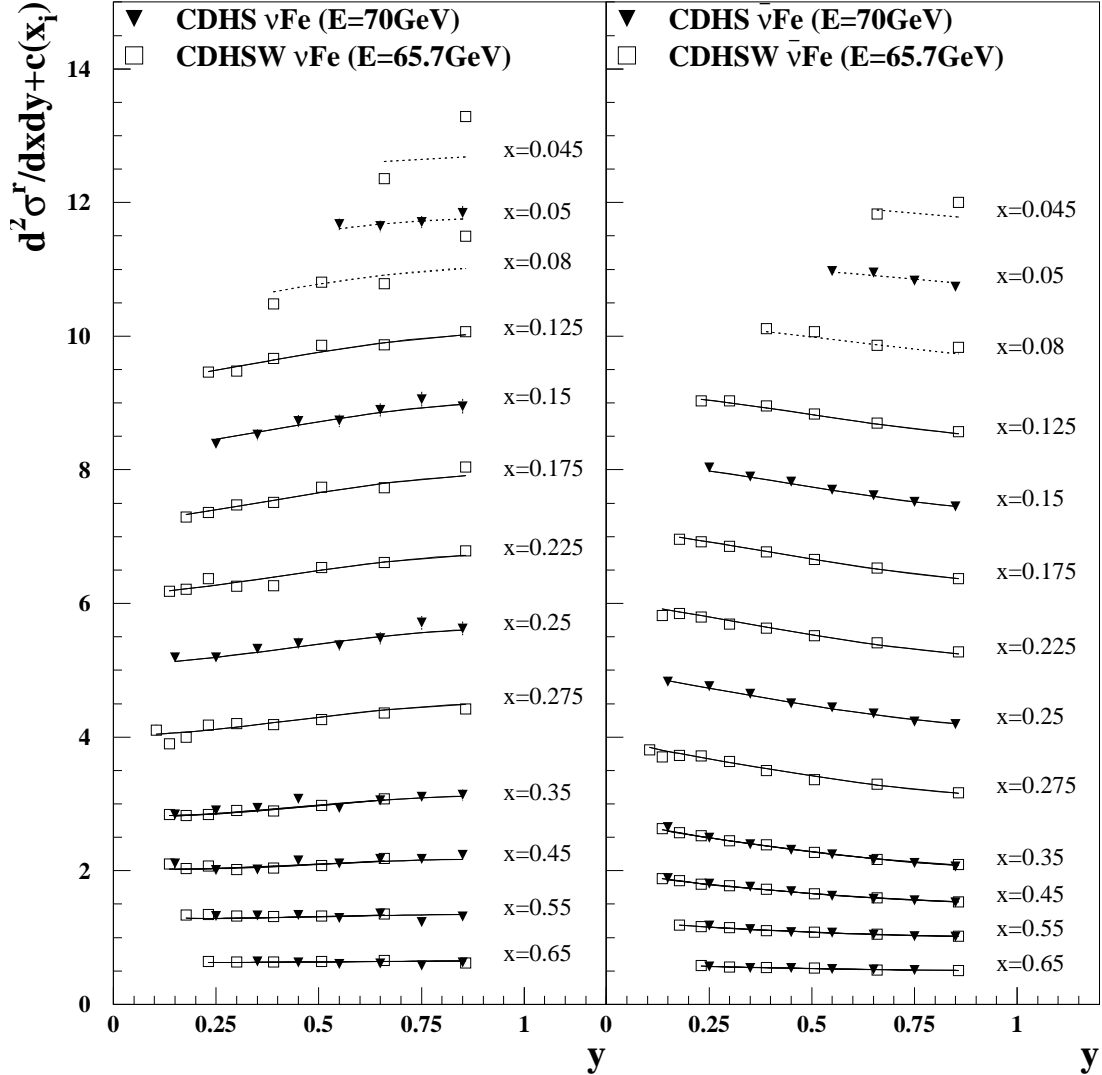}
    \caption{ Same as fig.~\ref{fig-cdhs} but for the CDHSW 65.7 
GeV beam data
 sample. The CDHS data ($E=70$ GeV), though not entering the fit, 
are also shown.
 The full and dotted 
lines show the {\tt fit1} results.
 The dotted lines describe the data rejected from
{\tt fit1} ($x<0.1$). The nuclear correction for these 
particular data are 
taken from
ref.~\cite{seligman}.
 The error bars represent the quadratic sum of
 statistical and uncorrelated systematic errors.
 The  additive bin constants are
 $c(x_i)=\{ 12, 11, 10, 9, 8, 7, 6, 5, 4, 3, 2, 1.5, 1, 0.5 \} $}.
    \label{fig-cdhsw}
  \end{center}
\end{figure}

 The reduced $\nu$ ($\bar{\nu}$)
 differential cross section
\begin{equation}\label{sigmared}
\frac{d^2\sigma^r}{dxdy}=\frac{2\pi(M_W^2+Q^2)^2}{G_F^2 M_N M_W^4E}
\frac{1}{Y}\frac{d^2\sigma}{dxdy}\,,\,{\rm with}\,\,
 Y=1+(1-y)^2-\frac{M_Nxy}{E}\, ,
\end{equation}
computed from {\tt fit1} is compared to 
the data  in 
fig.~\ref{fig-cdhs} (Hydrogen target), fig.~\ref{fig-bebc} 
(Deuterium) and fig.~\ref{fig-cdhsw}
(Iron). 
The BEBC and CDHS Hydrogen and Deuterium data are well 
fitted 
down to small $x$. These data 
 contribute to 
constrain the valence 
distributions, without being affected by nuclear 
effects. 
Fig.~\ref{fig-cdhsw} also shows that the CDHS Iron data, 
though not entering
 the fit, are well described by it. 
As for the CDHSW rejected data ($x < 0.1$), discrepancies 
with the fit appear only  
in the first bin $x=0.045$ of figs.~\ref{fig-cdhsw}.

Figs.~\ref{fig-cdhsw}, \ref{bcdms}, \ref{h1-data} and
\ref{nmc-data} and the $\chi^2$
 results of table~\ref{tab:data-in-the-fit} show explicitly that 
the $\nu$ and $\bar{\nu}$ Iron data 
 are compatible with the data on 
$F_2$ coming from NC charged--lepton 
DIS even in the region $0.045<x<0.2$.
By contrast, in 
\cite{seligman,ccfr-alphas} a sizeable discrepancy is found 
between the CCFR $F_2^{\nu}$ and the NMC  $F_2^{\mu}$. 
From the analysis
performed here, which is done -- we recall -- 
on neutrino {\em cross sections},  
no disagreement emerges between charged-lepton
 and neutrino DIS measurements.  
We checked that including in our analysis the 
CCFR structure functions the fit worsens. We also 
found that the $F_3$ data are much better
 described than the $F_2$ data. We do not give any 
quantitative information about the $\chi^2$ of this particular
 fit since we considered
the statistical errors only (the use of the total systematic
 uncertainties is not recommended by the CCFR Collaboration).
In conclusion, some incompatibility 
seems to exist between the CCFR {\em structure functions} 
and all other charged-lepton and neutrino DIS data. 
A conclusive word on this matter could come  
from the analysis of the CCFR {\em cross 
sections}, which are  unavailable at the moment. 

\begin{figure}[htbp]
  \begin{center}
    \leavevmode
    \includegraphics*[width=6in]{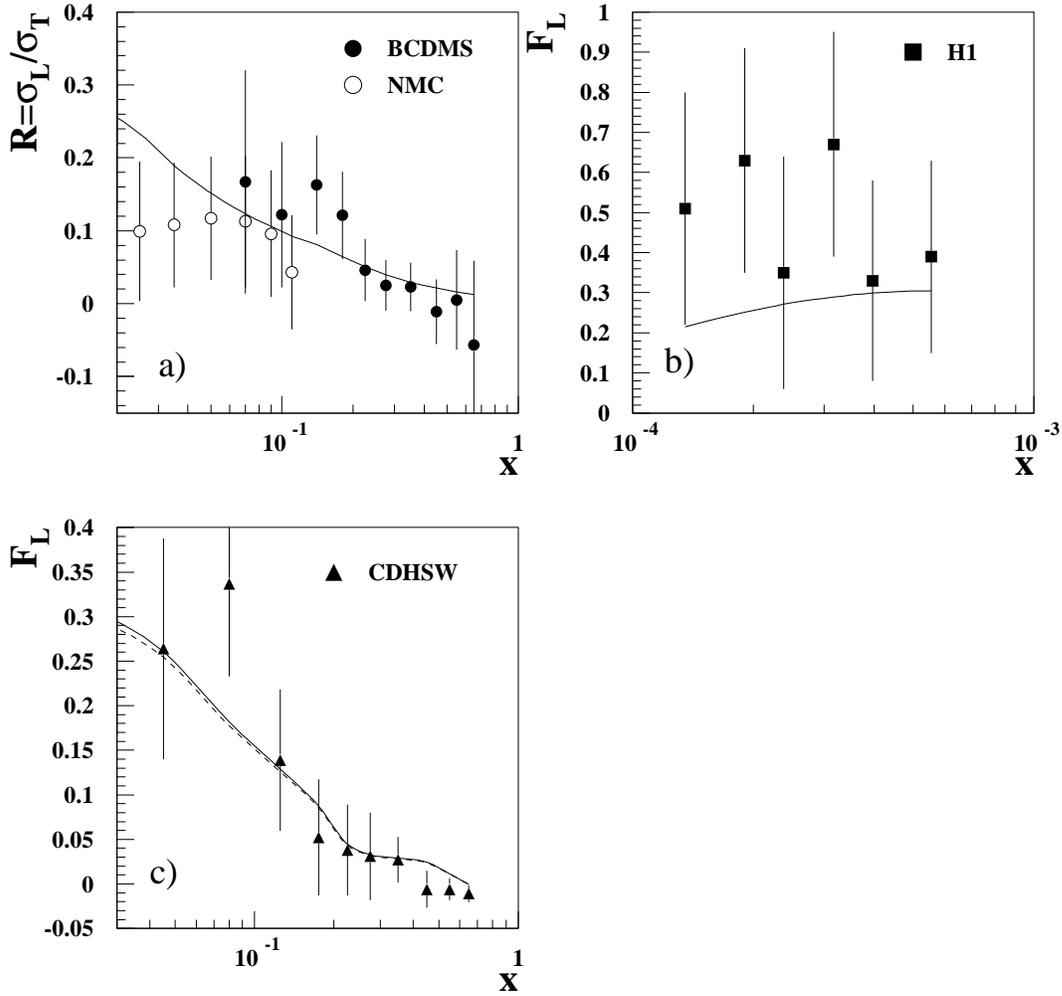}
    \caption{Results of {\tt fit1} 
for the longitudinal cross sections and structure 
functions. Comparison of $R$ and $F_L$ 
 calculated in {\tt fit1}  with: a) 
the BCDMS and NMC  measurements of $R= \sigma_L/\sigma_T$;
 b) the  H1  extraction of $F_L$; c) 
the CDHSW measurements (the dashed line is
$F_L^{\bar{\nu}}$, the solid line is $F_L^\nu$). 
None of these measurements
  enter our fits and only those corresponding to 
 $Q^2\ge3.5$GeV$^2$ are shown. The error bars 
represent the quadratic sum of
 the statistical and systematic errors.
 Each point of these plots corresponds to 
 different values of $Q^2$. The curves are obtained 
by interpolating smoothly 
 between the calculations performed at these points. }
    \label{fig-r}
  \end{center}
\end{figure}

The longitudinal structure function $F_L$ and 
the longitudinal to transverse cross section ratio $R$
\[
F_L=F_2-\biggl(1+\frac{4M_N^2x^2}{Q^2}\biggr)2xF_1\;,\;\;\;\;\;
R=\frac{F_2}{2xF_1}\biggl(1+\frac{4M_N^2x^2}{Q^2}\biggr)-1
\]
computed using the pdf's of {\tt fit1} (we remind 
that all our fits include target mass corrections) 
are compared to the NMC \cite{nmc} and BCDMS 
\cite{bcdms} results in 
fig.~\ref{fig-r}.a, to the H1 measurement \cite{h1-r} 
in fig.~\ref{fig-r}.b and to the CDHSW measurement \cite{cdhsw}
in fig.~\ref{fig-r}.c. Again, one can see a good agreement 
between our fit and the experimental results.

The beam energy dependence 
 of the CDHSW $\nu Fe$ and $\bar{\nu}Fe$ data, resulting from the 
global $\chi^2$ minimisation, is shown in fig.~\ref{fig-sigmae}.
The deviation from the independent linear fit of 
section~\ref{CDHSW-measurements} is small for $\bar{\nu}Fe$ data and larger
 for $\nu Fe$.

\subsection{The strange sea density}\label{fit2}

Let us concentrate now on the strange sea density. 
Due to the lack of data able to constrain it, this 
distribution plays a lesser role in the 
existing global fits. Two of them (CTEQ \cite{cteq} and 
MRST \cite{mrs}), guided by the results of the 
CCFR determination of $s(x)$ \cite{ccfr-strange}, in particular 
by the CCFR value of the strange-to-non-strange momentum ratio
$\kappa \equiv \langle x(s+\bar s) \rangle/
\langle x( \bar u + \bar d) \rangle \simeq 0.5$,  
impose
\begin{equation}
s(x) + \bar s(x) = \frac{1}{2} \, [\bar u(x) + \bar d(x)] \;. 
\label{sud}
\end{equation}

In the GRV analysis \cite{grv}, instead, the strange distribution 
is set to zero at the input scale, and then radiatively generated.   

The abundance of our neutrino and anti-neutrino data sets 
allows us to fit $s(x)$ with no extra constraints. 
The resulting 
 $s$  distribution of 
 {\tt fit1} is shown, with its error 
band, in fig.~\ref{pdf99}, where it is 
compared with the other fits. Notice that 
it turns out to be closer to MRST and CTEQ 
at large $x$, and to GRV at low $x$, where  
it is however much less constrained. 
From table~\ref{momentum-fraction} one sees that 
in {\tt fit1} $\kappa = 0.67$ at $Q^2 = 20$ GeV$^2$. 

\begin{figure}[htbp]
  \begin{center}
  \leavevmode
  \includegraphics*[width=6in]{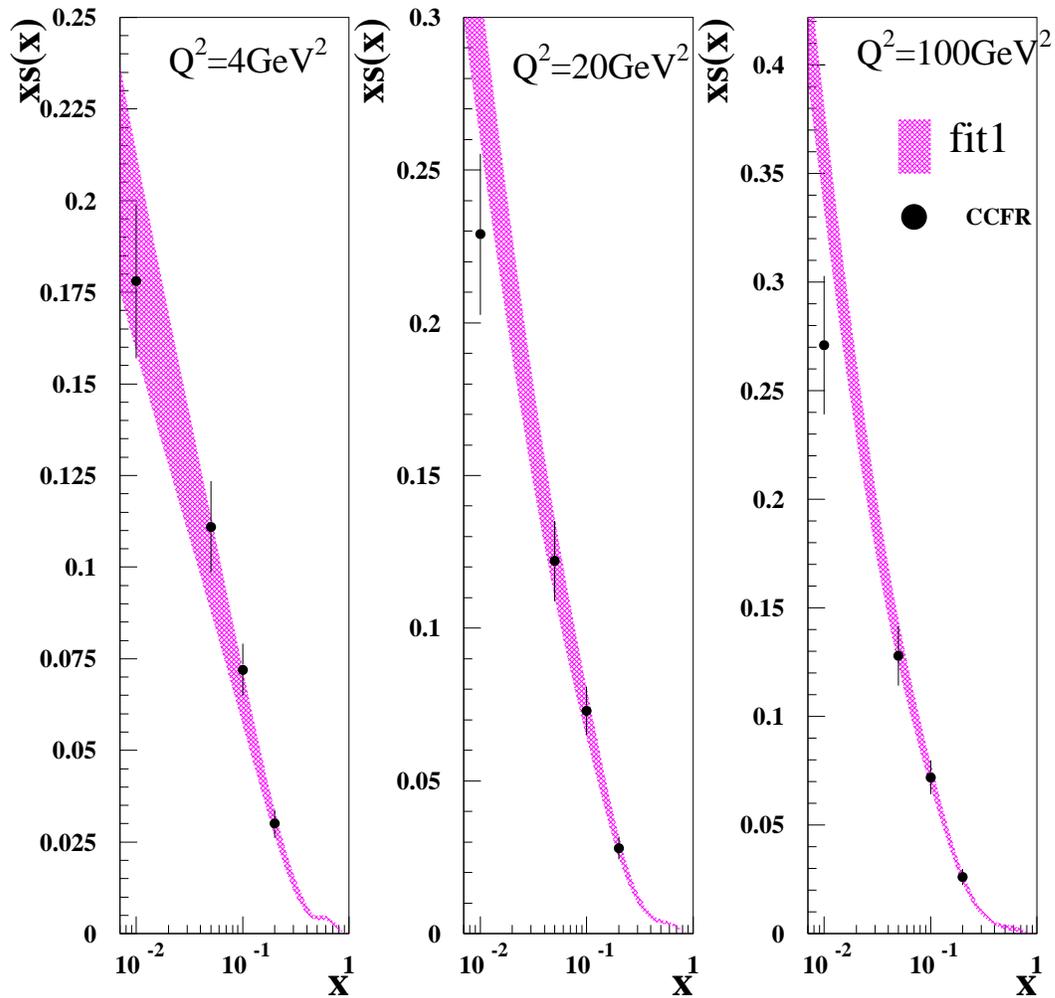}
  \caption{The strange distribution function  
of {\tt fit1}, with its error band, for three different 
values of $Q^2$. 
The CCFR dimuon determination (full circles with 
statistical and systematic errors added in quadrature) is also shown. }
  \label{s-plots}
  \end{center}
\end{figure}

In fig.~\ref{s-plots} 
$s(x)$ is plotted 
  at three different 
$Q^2$ values, together with 
the CCFR results \cite{ccfr-strange}. 
The agreement is good, 
 although  we must recall that 
the CCFR
extraction of the strange distribution from  dimuon 
production in $\nu(\bar \nu)Fe$ DIS has been criticised in many respects 
\cite{io,io2,gkr,gkr2}.

\begin{figure}[htbp]
  \begin{center}
    \leavevmode
    \includegraphics*[width=6in]{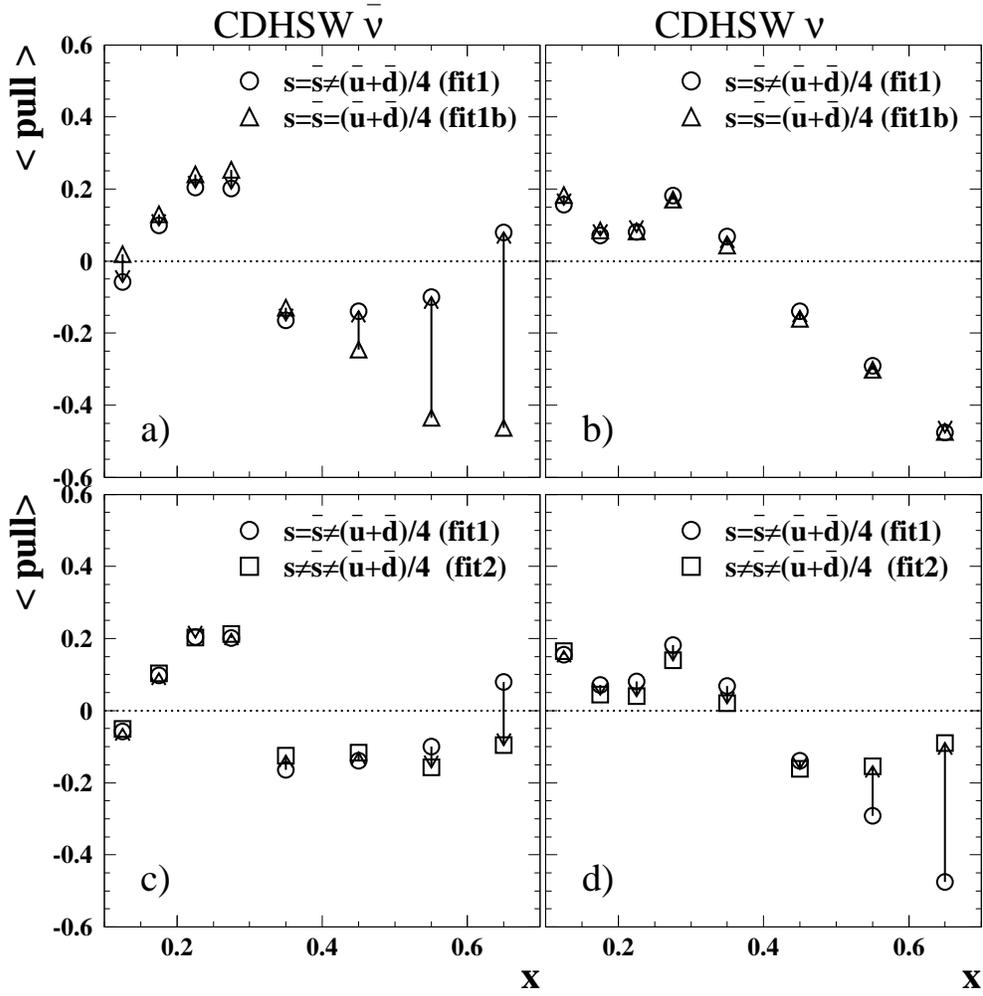}
    \caption{Mean value of the pull distribution as a function of 
$x$ (see text)
 for CDHSW anti-neutrino (a and c) and neutrino (b and d) data. Three fits
 are compared (see text). The arrows describe the changes of the pull 
values when 
 going from one fit to another.}
    \label{pull}
  \end{center}
\end{figure}

Fig.~\ref{pull} shows quantitatively how {\tt fit1} is 
favoured with respect to another fit, that we call 
{\tt fit1b}, in which the constraint (\ref{sud}) 
is imposed. The total $\chi^2$ of {\tt fit1b} is 2492.4, higher 
than the $\chi^2$ of {\tt fit1} by 62 units.
 In fig.~\ref{pull} we plotted the mean value of the so-called 
{\em pull}
 distribution,  
 as a function of $x$, for the CDHSW data: 
\begin{equation}
<pull>_{x_{dat}}=\frac{1}{N_{exp}(x_{dat})}
 \sum_{dat}
 \frac{{\cal O}^{dat}_{exp}-
{\cal O}^{ fit}\{1-\nu_{exp}\sigma_{exp}
-\sum_k\delta^{dat}_k(s^{exp}_k)\}}
{\sqrt{\sigma_{dat,stat}^2+ \sigma_{dat,uncor}^2}}\, ,
\end{equation}
where $N_{exp}(x_{dat})$ is the number of data at
 $x=x_{dat}$ for 
the experiment $exp$. In our case $exp$ stands for 
 CDHSW($\nu Fe$) or CDHSW($\bar{\nu} Fe$).
One can see from figs.~\ref{pull}.a and \ref{pull}.b that it is 
the $\bar \nu$ data which tend to favour {\tt fit1}
with respect to {\tt fit1b}. 

A very interesting question is whether the strange 
distribution is equal or not to the anti-strange one. 
The usual assumption $s(x) = \bar s(x)$ is  not 
dictated in fact by first principles.

We thus looked for a  
possible charge asymmetry of the strange sea  
performing a fit, called {\tt fit2}, in which we release the 
constraint $s = \bar s$. 
From table~\ref{table-neutrino-data}, 
one can see that the attempt of 
disentangling the $s$ and $\bar s$ distributions 
is justified by the abundance of anti-neutrino events in our data set.

The parametrisation (\ref{pdf_s}) for $s=\bar s$ is replaced 
in {\tt fit2} by two independent functions for $s$ and $\bar s$
\begin{eqnarray}
x s(x,Q_0^2) = 
A_{s} \, x^{B_{s}} \, 
(1-x)^{C_{s}}\, (1+D_{s} \, x^{E_{s}}),
\label{pdf_s2} \\
x \bar s(x,Q_0^2) = 
A_{\bar s} \, x^{B_{\bar s}} \, 
(1-x)^{C_{\bar s}}\, (1+D_{\bar s} \, x^{E_{\bar s}}).  
\label{pdf_sbar2}
\end{eqnarray}

We set $A_s = A_{\bar s}$ and $B_s = B_{\bar s}$, and 
we fix one more parameter by imposing 
$\int_0^1 (s - \bar s)\, dx =0$ (no net strangeness). 

\begin{figure}[htbp]
  \begin{center}
    \leavevmode
    \includegraphics*[width=6in]{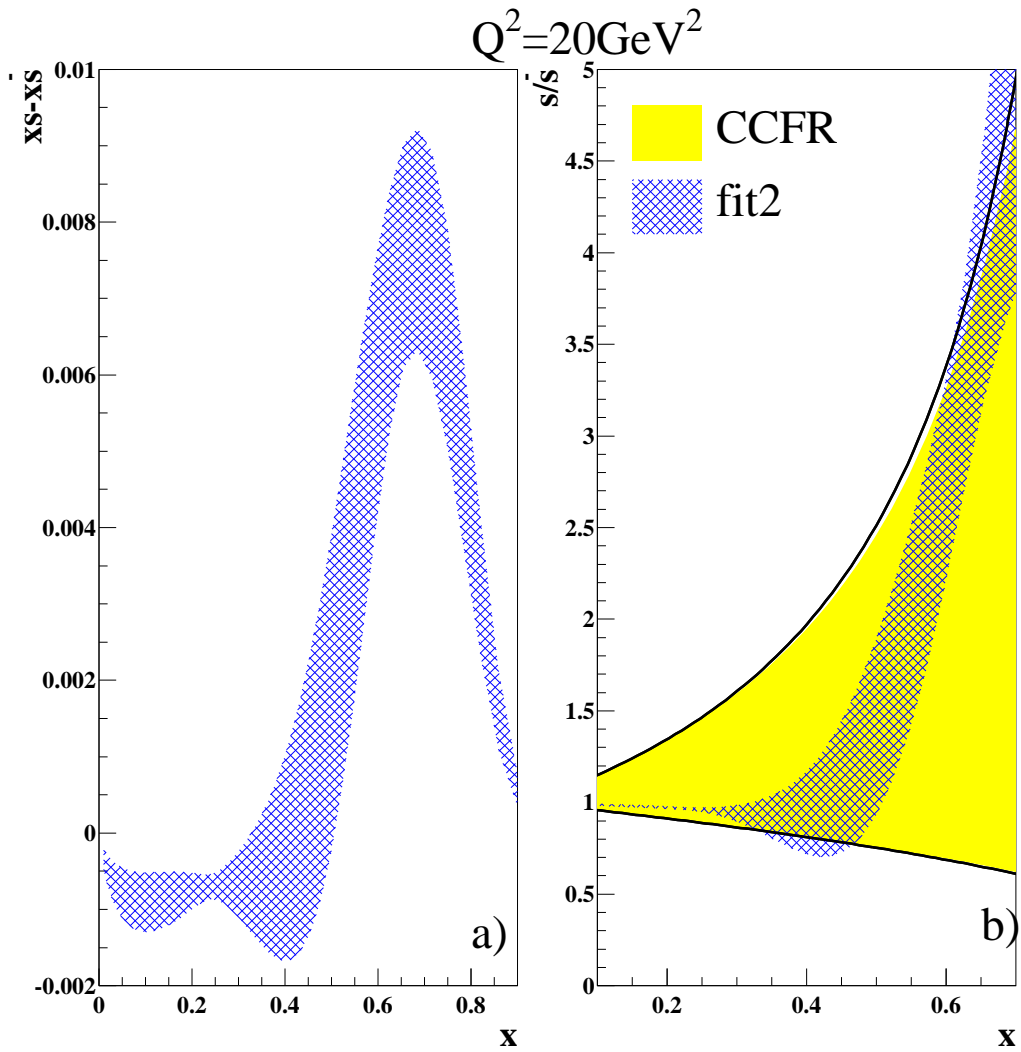}
    \caption{Results of {\tt fit2} for: a) the difference 
$x(s-\bar{s})$ and b) the ratio $s/\bar{s}$
 at $Q^2=20$GeV$^2$. In the box b) the result of CCFR is also shown. }
    \label{s_over_sbar}
  \end{center}
\end{figure}

The main results of {\tt fit2} are: 
\begin{itemize}
\item
The minimum $\chi^2$ decreases by 25 units with respect to {\tt fit1}  
(see table~\ref{tab:data-in-the-fit}). Hence the choice $s \neq \bar s$ 
is slightly favoured. 
\item
The strange distribution turns out 
to be  harder than the anti-strange one. 
The difference $s -\bar s$ is shown in 
fig.~\ref{s_over_sbar}.a with its error band. In fig.~\ref{s_over_sbar}.b 
we plot the ratio $s/\bar s$ at $Q^2 = 20$ GeV$^2$.  
\item
The momentum fractions of the {\tt fit2} partons 
at different $Q^2$ values are listed in table~\ref{momentum-fraction}. 
The momentum fraction $\langle x s \rangle$
is larger than   $\langle x {\bar s} \rangle$. 
Notice also that {\tt fit2} favours a higher value for the
 strange-to-non-strange momentum ratio $\kappa$, with respect 
to the fit with $s = \bar s$.  
\end{itemize}

Let us comment now on a previous test of the strange sea asymmetry. 
 In \cite{ccfr-strange} 
the CCFR Collaboration  found no 
evidence for $s \ne \bar s$. In their analysis 
the constraint 
$s(x)=[\bar{u}(x)+\bar{d}(x)]\times A(1-x)^C$ was imposed,
which limits the flexibility of the fit. 
Moreover, the dimuon sample 
of CCFR does not cover the high-$x$ region and consists only 
 of 5000 neutrino events and 1000 anti-neutrino events.  
Our analysis does not have  these limitations 
(our data sample is much more 
balanced between neutrino and anti-neutrino events and 
no extra constraints are set but those 
ensuring no net strangeness),
 and gives a more precise result on $s /\bar s$, as 
 one can see from fig.~\ref{s_over_sbar}.

\begin{figure}[b!]
  \begin{center}
    \leavevmode
    \includegraphics*[width=6in]{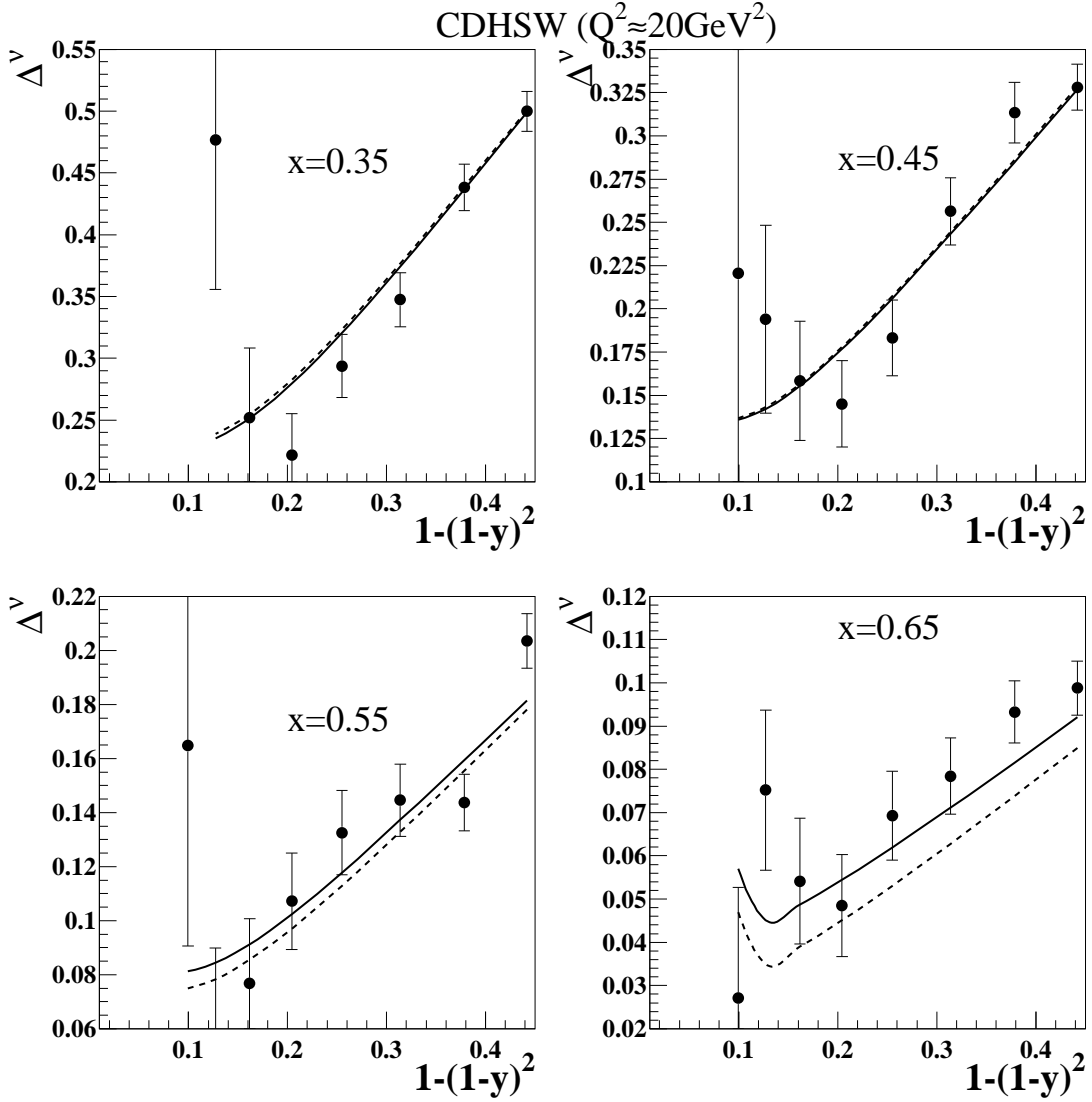}
    \caption{Difference between the $\nu Fe$ and 
 $\bar \nu Fe$ CDHSW differential cross-sections
 (see eq.~(\ref{delta_nu})) at fixed $x$ and $Q^2\approx 20$ GeV$^2$
 as a function of $Y_-=1-(1-y)^2$.
The solid line corresponds to {\tt fit2},
the dashed line to {\tt fit1}. The exact values of $Q^2$ are
 22, 21.6, 20.3, 18.6 GeV$^2$
for $x =  0.35,\, 0.45,\, 0.55,\, 0.65$, respectively.}
    \label{cdhsw_diff_18gev}
  \end{center}
\end{figure}

In order to understand how the difference $s-\bar s$ is constrained
by the data, let us consider the quantity
 \begin{equation}
   \label{delta_nu}
   \Delta^{\nu- \bar \nu}\equiv \frac{4\pi x(M_W^2+Q^2)^2}{G_F^2  M_W^4}
 \biggl[
\frac{d^2\sigma^{\nu N}}{dxdQ^2}-\frac{d^2\sigma^{\bar{\nu} N}}{dxdQ^2}
\biggr]\,.
 \end{equation}
The flavour content of $\Delta^{\nu - \bar \nu}$ is more evident 
in the parton model where it reads 
 \begin{equation}
   \label{delta_nu_qpm}
   \Delta^{\nu- \bar \nu} 
\propto x s(x) - x \bar{s}(x)+Y_-[xu_v(x)+xd_v(x)]
\,,
 \end{equation}
 with $Y_-=1-(1-y)^2$. The $\nu - \bar \nu$ cross section
 difference (\ref{delta_nu}) is plotted as a function of $Y_-$, at
 fixed $x$ and $Q^2$, in fig.~\ref{cdhsw_diff_18gev}.  
Comparing {\tt fit1} and {\tt fit2}, one can see that their
results deviate with increasing $x$ (being very close to each other 
 for $x\lesssim 0.3$). At high-$x$ 
{\tt fit1} undershoots the CDHSW values 
of $\Delta^\nu$  for all $Q^2$ bins. 
Fig.~\ref{cdhsw_diff_18gev} shows that the 
 CDHSW data are more precise at high-$y$. Hence it is
this region which drives the result on $s - \bar s$. 
Looking at fig.~\ref{pull}.c,d  one sees also that it is 
the $\nu$ data which favour {\tt fit2} with respect to {\tt fit1}. 

\begin{figure}[htbp]
  \begin{center}
    \leavevmode
    \includegraphics*[width=6in]{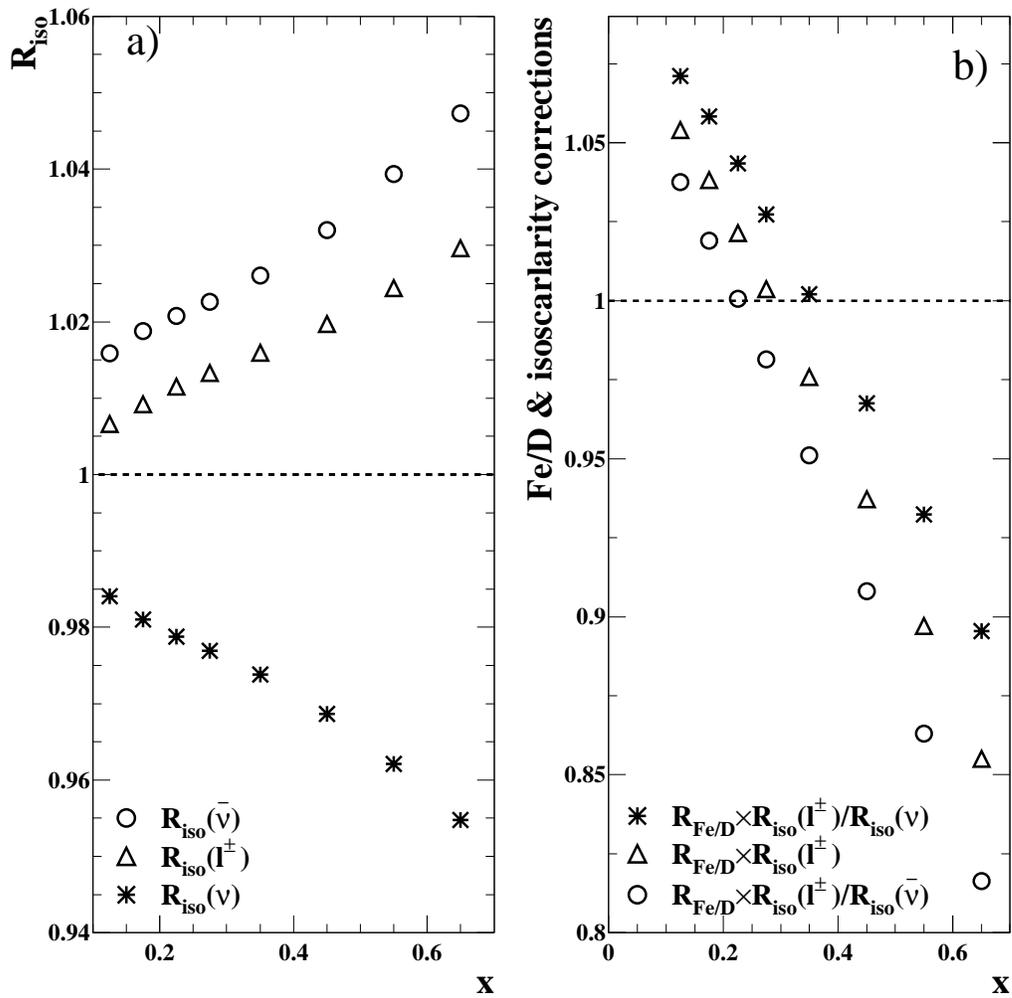}
    \caption{ Corrections applied to Iron target data: 
 a) isoscalarity corrections;  b) nuclear and isoscalarity
   corrections.}
    \label{isocor}
  \end{center}
\end{figure}

Since the nuclear corrections 
applied to the CDHSW data 
are sizable (see fig.~\ref{isocor}.b), 
one may naturally ask to what 
extent our results are affected 
by the uncertainties on the evaluation of these corrections. 
The $Fe/D$ ratio, as we have seen 
(see fig.~\ref{fig-iron/D} and the 
 check described at the end of section~\ref{procedure}),
is well determined 
by the charged lepton data. Moreover, 
 it factorises in the $\nu - \bar \nu$ difference. 
Hence this component of 
the  nuclear correction is rather harmless. 
The isoscalarity corrections are instead 
different for neutrino and 
anti-neutrino observables and quite large. 
Looking 
 at eqs.~(\ref{ison}, \ref{isoscalar})
one sees that the isoscalarity
ratio
 cannot exceed $A/(2Z)=1.073$ for Iron. 
Fig.~\ref{isocor}.a shows that 
the maximal value we obtained is not far from this 
upper bound. 
We checked that 
reducing the isoscalarity ratio the resulting 
$s- \bar s$ difference gets larger. Thus
 we conclude that
our results on the strange sea asymmetry are not spoiled, 
at least qualitatively, by the 
uncertainties on the isoscalarity corrections.

Theoretically, a charge asymmetric sea is accounted for by 
introducing a distinction between extrinsic and intrinsic 
$q \bar q$ pairs \cite{brodsky}. 
The extrinsic sea consists of short--lived 
quarks and anti--quarks produced by QCD hard subprocesses 
(bremsstrahlung and gluon splitting). It is evident that 
the extrinsic component of the sea cannot be charge asymmetric. 
On the other hand, the intrinsic $q \bar q$ pairs exist 
over a longer time scale and are associated with nonperturbative
phenomena. These pairs are still produced by gluon 
fragmentation but have time, before recombining,
 to interact with other partons. They 
represent higher Fock states 
of the nucleon 
($\vert qqq q \bar q \ldots 
\rangle$) 
and manifest themselves in meson--baryon 
fluctuations. 

There is no fundamental principle forbidding
a possible charge asymmetry of the intrinsic sea. 
Actually, there are reasons to believe that such asymmetry
should indeed be a property of the strange and 
charmed sea \cite{brodsky,burkardt}. 
If the strange (or charmed) sea is asymmetric  
at low $Q^2$ due to  some nonperturbative mechanism, 
the QCD evolution simply preserves this asymmetry because 
$s- \bar s$ (or $c - \bar c$) evolves like a non-singlet
distribution and its first moment is constant. 
An interesting feature of the intrinsic sea is that 
it tends to exist at relatively 
large values of $x$ \cite{brodsky,burkardt}, 
corresponding to the most energetically favoured 
configuration of the nucleon light--cone wave function.  

In the simplest model \cite{thomas,burkardt} 
the production of the intrinsic strange sea is attributed 
to the $p \rightarrow \Lambda K^{+}$ fluctuation. Due 
to chiral symmetry pseudoscalar mesons 
have relatively small masses. 
 As a consequence 
\cite{burkardt}, the average $x$ of the 
$\bar s$ anti-quark in the $K$ is smaller that 
the average $x$ of the $s$ quark coming from the 
$\Lambda$. Thus the $s$ distribution is 
expected, on quite general grounds, to be 
harder than the $\bar s$ distribution. This expectation 
has been substantiated by explicit 
calculations in \cite{thomas,ma} (for other models see  
\cite{ji}). Our results on the strange
and anti-strange sea density 
are, at least qualitatively, in agreement 
with the predictions of the intrinsic sea theory.

\section{Conclusions}

We have presented a global next-to-leading 
order QCD analysis of a large set of DIS data, 
including the (properly re-evaluated) neutrino and 
anti-neutrino differential cross sections of BEBC, CDHS, CDHSW, 
 the charged-lepton structure functions of  
 NMC, BCDMS and H1, and the Drell-Yan data of E605, NA51 and 
E866.
The full use of the information on the nucleon structure 
embodied in neutrino DIS observables and a proper treatment
 of the experimental systematic uncertainties are the main novelties 
of our approach. 
 In particular, the large-statistics 
CDHSW Iron data 
allow disentangling the 
strange sector from the non strange one: this leads 
to a consistent determination of $s(x)$ within 
a global fit, similarly to what happens for 
 the other parton distributions.  

The charm mass effects, whose relevance for an accurate 
determination of the parton densities is a recent firm 
acquisition, have 
been consistently treated in a massive factorisation 
scheme, the Fixed Flavour Scheme. 

We found no evidence of any discrepancy between the 
 neutrino data we considered in our fit 
and the charged--lepton data. 
A complete and unambiguous analysis of {\it all}
neutrino experiments and a conclusive check of the
compatibility of their data with the charged-lepton data
would be possible only if we could use the CCFR differential cross
sections, which are unfortunately not available. 

The large statistics of anti-neutrino events in our data sample
allowed us to test the hypothesis 
of a charge asymmetric strange sea: $s \ne \bar s$. 
We found some evidence for such an asymmetry. 
The qualitative features of the resulting $s$ and $\bar s$
distributions (namely, a large $x$ tail at low $Q^2$ 
 and $\bar s(x)$ softer than $s(x)$) agree  
with the expectations of the intrinsic sea theory.

Finally, we outline some developments. 
 As for the strong coupling, in the present work we 
took a pragmatic attitude, using 
the world average value. 
 A more systematic study, in which we will 
 extract $\alpha_s$ 
from the DIS data and investigate its correlation 
with the gluon density, will be presented in a  
forthcoming paper. 

Another important development 
that we have in mind is 
the comparison of the results obtained in the two 
QCD massive schemes, FFS and VFS. 

Finally, one should envisage some independent, and perhaps more
direct, experimental test of 
the charge asymmetry of the strange sea for which we 
presented here only an indirect statistical evidence.

\section*{Acknowledgments}

 We thank J.~Bl{\"u}mlein, V.~Del Duca, S.~Forte,
G.~Korchemski,
  M.W.~Krasny, B.-Q.~Ma, V.~Massoud,
 W.~Melnitchouk, A.~Milsztajn, N.N.~Nikolaev 
and W.~Seligman for valuable help
 and/or useful 
 discussions. Two of us
 (C.P. and F.Z.) would like to thank their H1 colleagues for their help 
in elaborating the pQCD analysis program.

\section*{Appendix I: Unfolding procedure}

 A two dimension net in $x$ and $y$ is defined. The nodes correspond to the
centres of the experimental bins.
 Following \cite{nous-method}, we define a 
basis of continuous functions $\phi_i(x)$ and  $\psi_j(y)$ such 
that $\phi_i(x^c_l)=\delta_{il}$ and $\psi_j(y^c_m)=\delta_{jm}$ 
(where $\delta_{il}$ is the Kronecker symbol). Provided these functions
are determined using a spline interpolation \cite{spline}, the
 differential cross section  can be written as
\begin{equation}
  \label{spline1}
\frac{d^2\sigma^{\nu(\bar{\nu})Fe}(x,y,E^c_k)}{dxdy}=
\sum_{l=1}^{n_x}\sum_{m=1}^{n_y}
\phi_i(x)\psi_j(y)
\frac{d^2\sigma^{\nu(\bar{\nu})Fe}(x_l^c,y_m^c,E^c_k)}{dxdy}.
\end{equation}
Applying the average theorem, eq.~(\ref{spline1}) leads to
\begin{equation}\label{system-1}
\sigma_{ijk}
=\sum_{l=1}^{n_x}\sum_{m=1}^{n_y}c_x^{i,l}c_y^{j,m}
\frac{d^2\sigma^{\nu(\bar{\nu})Fe}(x_l^c,y_m^c,E^c_k)}{dxdy}
\end{equation}
with
\[
\sigma_{ijk}\approx \frac{N_{ijk}}{C},\quad
c_x^{i,l}=\frac{\int_{x_i}^{x_{i+1}}\phi_l(x)dx}{x_{i+1}-x_i},\quad
c_y^{j,m}=\frac{\int_{y_j}^{y_{j+1}}\psi_m(y)dy}{y_{j+1}-y_j}
\]
where $N_{ijk}$ is the number of events experimentally 
observed and $C$ is 
the number of scattering centres in the target.

The differential cross sections at the bin centres 
$d^2\sigma^{\nu(\bar{\nu})Fe}(x_l^c,y_m^c,E^c_k)/dxdy$ are obtained by
 reversing this system of $n_x\times n_y$ equations.
 Numerically, the more stable results were obtained using the first order
 spline.
 In this case, $\phi_i(x)$ and  $\psi_j(y)$ are the Lagrange 
(or `hat') functions \cite{spline} 
\begin{equation}\label{phi}
\phi_i(x)=
\begin{cases}
(x-x_{i-1})/(x_i-x_{i-1}) & \hbox{,   }x_{i-1}\le x \le x_i \\
(x_{i+1}-x)/(x_{i+1}-x_i) & \hbox{,   }x_i\le x \le x_{i+1} \\
0                         & \hbox{,   otherwise}.
\end{cases}
\end{equation}
The same expression holds for $\psi_j(y)$.

However,  using eq.~(\ref{phi}), one can see that the system of 
eq.~(\ref{system-1}) is incomplete if $x_{1}\ne 0$ or
 $x_{n_x}\ne 1$ or $y_{1}\ne 0$  or $y_{n_y}\ne 1$. Indeed, the 
 boundary bins receive some contributions 
(in the r.h.s. of eq.~(\ref{system-1})) coming from 
 the bin neighbours which are not included in the measurements. 
  To overcome this difficulty, we have defined artificial
extra bins $[x_{n_x},x_{n_x+1}]$ $\forall y$, $[y_{0},y_{1}]$ $\forall x$ and
 $[y_{n_y},y_{n_y+1}]$ $\forall x$. Then we fixed  
 the cross section at the centre of these new bins to a
certain fraction $\lambda$ of 
the closest cross section measurement. Under this modification,
 eq.~(\ref{system-1}) becomes an inhomogeneous system
 \begin{eqnarray}\label{system-2}
\sigma_{i,j,k}-\lambda\biggl(c_x^{i,n_x+1}c_y^{j,0}\sigma_{n_x,1,k}
+c_x^{i,n_x+1}c_y^{j,n_y+1}\sigma_{n_x,n_y,k}+\nonumber\\
\sum_{m=1}^{n_y} c_x^{i,n_x+1}c_y^{j,m}\sigma_{n_x,j,k}
 +
\sum_{l=1}^{n_x}
c_x^{i,l}[c_y^{j,0}\sigma_{l,1,k}+c_y^{j,n_y+1}\sigma_{l,n_y,k}] 
\biggr)=\nonumber\\
\sum_{l=1}^{n_x}\sum_{m=1}^{n_y}c_x^{i,l}c_y^{j,m}
\frac{d^2\sigma^{\nu(\bar{\nu})Fe}(x_l^c,y_m^c,E^c_k)}{dxdy}
\end{eqnarray}
and the differential cross sections at the bin centres are obtained by
inverting this system with $\lambda=1$.

 In order to test the sensitivity of our results
to the choice of $\lambda$, we have set $\lambda$ to some extreme 
values:
  $\lambda=2$, $\lambda=1/2$. We observed a variation of the order of 
$\approx 20\%$ in the highest $x$ bin for all $y$ bins. But we point out
 that
the results are completely stable in the first $x<0.1$ bins for 
all $y$ bins.
  This method is then used
 to perform the definitive bin centre correction in these 
particular bins, 
 as mentioned in section \ref{CDHSW-measurements}.
 The uncertainty of the method is estimated by repeating the procedure
 using a second order spline interpolation to define the
 basis functions $\phi_i(x)$ and  $\psi_j(y)$. The ratio of the two bin 
centre correction factors is then taken, bin by bin,
 as an estimate of the uncertainty due to 
 the method, and  added in quadrature to the uncorrelated 
systematic error
 of the measurement (it is of the order of $\lesssim 2\%$ and reaches 
$7.5\%$ at high $y$).

\section*{Appendix II: Error bands}

We describe here the formula used to calculate the error bands 
shown throughout the paper. 
Calling $\bmp \equiv \{p_1,...,p_n\}$ the vector of 
the free parameters
 of the fit, the error
 band of a given function $f$ is given at  each
$(x,Q^2)$ point by \cite{nous-error}:
\begin{equation}\label{band1}
\Delta f(x,Q^2;\bmp_0)
= \vert f(x,Q^2;\bmp_0 + \bmd_p(x,Q^2))-
f(x,Q^2;\bmp_0- \bmd_p(x,Q^2)) \vert
\end{equation}
where $\bmp_0$
denotes the parameter set minimising the $\chi^2$ and
the vector \\ $\bmd_p(x,Q^2)
\equiv \{\Delta_{p_1}(x,Q^2),...,\Delta_{p_n}(x,Q^2)\}$
 is given by
\begin{equation}\label{band2}
\bmd_{p}(x,Q^2)
= \frac{M^{-1}\bmpart_p f(x,Q^2;\bmp)}{\sqrt{\bmpart_p 
f(x,Q^2; \bmp) M^{-1}
 \bmpart_p f(x,Q^2; \bmp)}}
\end{equation}
with $M_{ij}=(1/2) \,\partial^2\chi^2/\partial p_i\partial p_j$ and
$ \bmpart_p=\{\partial/
\partial p_1,...,\partial/\partial p_n\}$. 


\newpage

\begin{figure}[htbp]
  \begin{center}
    \leavevmode
    \includegraphics*[width=6in]{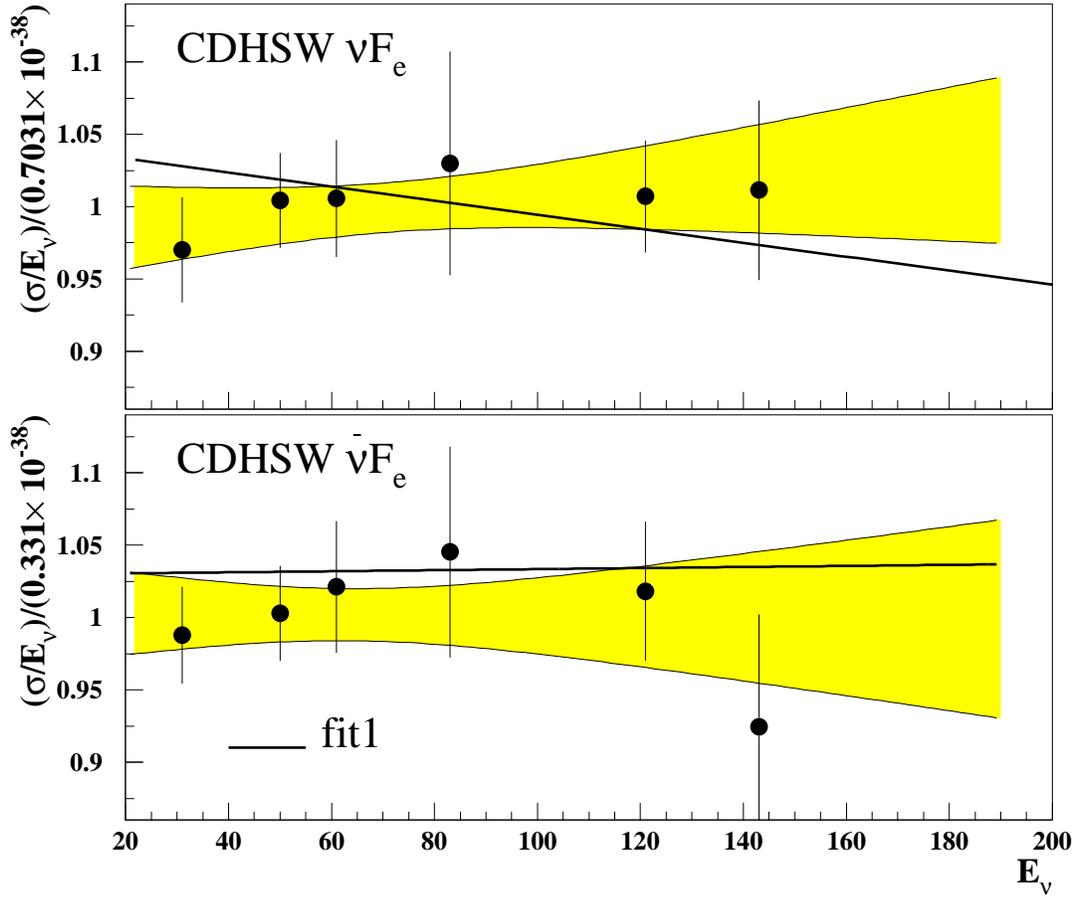}
    \caption{Total cross sections  of $\nu Fe$ (top)
 and $\bar{\nu} Fe$ (bottom) from CDHSW. The shaded areas are
the one-standard-deviation error bands corresponding 
to the linear fits described
 in section \ref{CDHSW-measurements}. The curves are
the results of  
{\tt fit1}.
 The error bars correspond to the quadratic sum of statistical
 and systematic errors.}
    \label{fig-sigmae}
  \end{center}
\end{figure}

\newpage

\end{document}